\documentclass{aa}
\usepackage{natbib}
\usepackage{xspace}
\usepackage{amssymb}
\usepackage{amsmath}
\usepackage{hyperref}
\usepackage{graphicx}

\def\revised#1{{#1}}
\def\revisedd#1{{#1}}

\def\rcyl{\ensuremath{r_\mathrm{cyl}}\xspace}
\def\hpgas{\ensuremath{h_\mathrm{p}}\xspace}
\def\tmid{\ensuremath{T_\mathrm{mid}}\xspace}

\def\smalldust{\ensuremath{\mathrm{small}}\xspace}
\def\bigdust{\ensuremath{\mathrm{big}}\xspace}
\def\dustgas{\ensuremath{\mathrm{dg}}\xspace}
\def\Qlambda{\ensuremath{Q_\lambda}\xspace}
\def\St{\ensuremath{\mathrm{St}}\xspace}
\def\coag{\ensuremath{\mathrm{X}}\xspace}
\def\comma{\,,}
\def\fullstop{\,.}
\begin{document}
\title{Razor-thin dust layers in protoplanetary disks:\\ Limits on the vertical shear instability}
\titlerunning{Razor-thin dust layers: Limits on the vertical shear instability}
\authorrunning{Dullemond, Ziampras, Ostertag, Dominik}
\author{C.P.~Dullemond$^{1}$, A.~Ziampras$^{2}$, D.~Ostertag$^{1}$, C.~Dominik$^{3}$}
\institute{(1) Institute for Theoretical Astrophysics, Center
  for Astronomie (ZAH), Heidelberg University, Albert Ueberle Str.\ 2,
  69120 Heidelberg, Germany; Email: dullemond@uni-heidelberg.de\\
  (2) Astronomy Unit, School of Physics and Astronomy, Queen Mary University of London, London E1 4NS, UK\\
  (3) Anton Pannekoek Institute for Astronomy, University of Amsterdam, Science Park 904, 1098XH Amsterdam, The Netherlands
}
\date{\today}

\abstract{
  {\em Context:} Recent observations with the Atacama Large Millimeter Array (ALMA)
  have shown that the large dust
  aggregates observed at millimeter wavelengths settle to the midplane into a
  remarkably thin layer. This sets strong limits on the strength of the
  turbulence and other gas motions in these disks.\\
  {\em Aims:} We intend to find out if the geometric thinness of these layers is
  evidence against the vertical shear instability (VSI) operating in these
  disks.  We aim to verify if a dust layer consisting of large enough dust
  aggregates could remain geometrically thin enough to be consistent with the
  latest observations of these dust layers, even if the disk is unstable
  to the VSI. If this is falsified, then the observed flatness of these dust
  layers proves that these disks are stable against the VSI, even out to the
  large radii at which these dust layers are observed.\\
  {\em Methods:} We performed hydrodynamic simulations of a protoplanetary disk
  with a locally isothermal equation of state, and let the VSI fully develop.
  We sprinkled dust particles with a given grain size at random positions near
  the midplane and followed their motion as they got stirred up by the VSI,
  assuming no feedback onto the gas. We repeated the experiment for different
  grain sizes and determined for which grain size the layer becomes thin enough
  to be consistent with ALMA observations. We then verified if, with these grain
  sizes, it is still possible (given the constraints of dust opacity and
  gravitational stability) to generate a \revised{moderately} optically thick layer at millimeter
  wavelengths, as observations appear to indicate.\\
  {\em Results:} We found that even very large dust aggregates with Stokes
  numbers close to unity get stirred up to relatively large heights above the
  midplane by the VSI, which is in conflict with the observed geometric thinness.  For
  grains so large that the Stokes number exceeds unity, the layer can be
  made to remain thin, but we show that it is hard to make dust layers optically
  thick at ALMA wavelengths (e.g.,~$\tau_{\mathrm{1.3mm}}\gtrsim 1$) with such large
  dust aggregates.\\
  {\em Conclusions:} We conclude that protoplanetary disks with geometrically
  thin midplane dust layers cannot be VSI unstable, at least not down to
  the disk midplane. Explanations for the
  inhibition of the VSI out to several hundreds of au include a high dust-to-gas
  ratio of the midplane layer,
  \revised{a modest background turbulence,} and/or a reduced
  dust-to-gas ratio of the small dust grains that are responsible for the
  radiative cooling of the disk. \revised{A reduction of small grains by a factor
    of between \revisedd{10 and 100} is sufficient to quench the VSI. Such a reduction
    is plausible in dust growth models, and still consistent with observations
  at optical and infrared wavelengths.}
}

\maketitle

\begin{keywords}
accretion, accretion disks -- protoplanetary disks
\end{keywords}

\section{Introduction}
According to canonical theory, the evolution of protoplanetary disks is thought
to be driven by a combination of viscous evolution and photoevaporation
\citep[e.g.,][]{2001MNRAS.328..485C}. The viscosity in such disks is thought to
be caused by turbulence produced by the magnetorotational instability
(MRI). While the MRI is inhibited in very dense regions of the disk (the
so-called dead zones), it may still be operational in the hot inner regions
($r\ll 1\,\mathrm{au}$), in the irradiated surface layers, and in the weakly
ionized outer regions ($r\gg 1\,\mathrm{au}$) of the disk
\citep[e.g.,][]{2013ApJ...765..114D}.

However, in recent years the turbulent viscous disk theory for the outer regions
of protoplanetary disks has been called into question. Using radiative
transfer modeling of the Atacama Large Millimeter Array (ALMA) image of HL Tau, \citet{2016ApJ...816...25P} infer
that the turbulence in that disk must be weak ($\alpha\lesssim 10^{-3}$). Direct
measurements of the turbulent line width in CO 2-1 with ALMA show mostly upper
limits to the turbulent velocities of $\lesssim 10\%$ of the local sound speed
\citep[e.g.,][]{2020ApJ...895..109F}. While these velocity upper limits are still
consistent with turbulent $\alpha$ values of up to $10^{-2}$, and marginally
consistent with MRI-turbulent disks \citep{2017ApJ...850..131F}, these and other
measurements have stimulated the exploration of the possible consequences of the
absence of MRI turbulence in protoplanetary disks and, consequently, the
possibility that these disks may be much less turbulent than previously thought.

The implications of very low turbulent $\alpha$ values for protoplanetary disks
are numerous. For instance, \citet{2017ApJ...850..201B} show that in
low-$\alpha$ disks a single planet can produce multiple rings. Indeed,
\citet{2018ApJ...869L..47Z} demonstrate that a protoplanetary disk model with
very low $\alpha$ and a single embedded planet can reproduce the
observed many-ringed structure of the disk around AS 209 remarkably well \citep[see,
  however][]{2020A&A...637A..50Z}. Low turbulent velocities also have strong
implications for dust growth, gap formation, planet migration, and
many other things. The reason why, until not long ago, turbulent $\alpha$
values of $\lesssim 10^{-3}$ were not seriously considered by scientists in the
field is that most stars with protoplanetary disks are observed to undergo
substantial gas accretion. This requires $\alpha$ values in excess of about
$10^{-3}$ \citep[e.g.][]{1998ApJ...495..385H}. However, wind-driven accretion
may provide a solution to this dilemma \citep[e.g.,][]{1995A&A...295..807F,
  2021MNRAS.tmpL.109T,2022arXiv220502126M}.

In the absense of the MRI, a protoplanetary disk can be prone to other,
non-magnetohydrodynamic instabilities that cause turbulence or turbulence-like
velocity fluctuations \citep[e.g.,][]{2019ApJ...871..150P}. Of particular
importance in the outer regions of the protoplanetary disk is the vertical shear
instability \citep[VSI, e.g.][]{2013MNRAS.435.2610N, 2014A&A...572A..77S,
  2016A&A...594A..57S,2020A&A...644A..50F}. This instability, when fully
developed, produces upward and downward vertical streams of gas that slowly
oscillate. \revised{These oscillations form a radially propagating wave
  \citep{10.1093/mnras/stac1598}.} When viewed as a kind of turbulence, it is
highly anisotropic, with turbulent ``eddies'' being radially narrow, but
vertically extended sheets of gas moving either up or down.  This ``turbulence''
is only weakly effective as a replacement of MRI turbulence for the radial
transport of angular momentum, with values on the order of
$\alpha_{\mathrm{VSI,radial}}\sim 10^{-4}$ \citep{2014A&A...572A..77S}. However,
due to the strong upward and downward motions of the gas, crossing the midplane,
with velocities in the range of 5\% to 20\% of the isothermal sound speed, the
effect of the VSI on the dust population of the disk is very pronounced
\citep{2016A&A...594A..57S, 2017ApJ...850..131F, 2019MNRAS.485.5221L,
  2022A&A...658A.156L}. Even
large dust particles can be stirred up to high elevations above the midplane.

From the observational side, however, there is now increasing evidence that
many, if not most, protoplanetary disks contain a layer of large dust aggregates
at the midplane that contains a substantial amount of dust mass and is
geometrically extremely flat, i.e., having a very small scale height. The first
evidence came from the ALMA image of the disk around HL Tau, where radiative
transfer modeling put an upper limit on the vertical scale height of the dust
layer of 1 au at a radius of 100 au \citep{2016ApJ...816...25P}. The high
resolution ALMA images of the DSHARP campaign \citep{2018ApJ...869L..41A} also
suggest very flat geometries of the dust layers seen at
$\lambda=1.3\,\mathrm{mm}$ wavelengths. Detailed radiative transfer analysis of
the DSHARP observations of HD 163296 shows that the dust in the inner ring of
that source (at $r\simeq 67\,\mathrm{au}$) appears to be vertically extended
almost to the gas pressure scale height, but the outer ring (at $r\simeq
100\,\mathrm{au}$) appears to be less than 10\% of the gas pressure scale
height, i.e., highly settled \citep{2021ApJ...912..164D}.

To get better constraints on the vertical extent (geometric thickness) of the
midplane dust layers of protoplanetary disks, dedicated observing campaigns with
ALMA for nearly-edge-on disks are required. The first such campaign already
yielded indications of strong settling of large grains
\citep{2020A&A...642A.164V}. But when the disk of Oph 163131 was reobserved
with ALMA by \citet{2022arXiv220400640V}, the vertical scale height of the dust
layer could be constrained to be less than 0.5 au at a radius of 100 au,
which is about 7\% of the gas pressure scale height at that radius. From
these observations, and under some assumptions of the grain size, these
authors derive an upper limit of $\alpha\lesssim 10^{-5}$ on the turbulence.

However, if the VSI is operational in these outer disk regions, one might
expect that the dust layer should be much more vertically extended, due to
the high efficiency of the vertical dust stirring of the VSI.
The purpose of this paper is to quantify this.

The grain sizes are not perfectly known, nor is the gas disk density. We address
the question whether the grains could be so large that they remain in a thin
layer in spite of the VSI. And if not, what could be the reason that the VSI is
not operational in this disk.


The paper is structured as follows: We start with an analysis of the stirring-up
of particles in Section \ref{sec-stirring}. In Section \ref{section-case-against-St-big}
we explain why $\mathrm{St}\gg 1$ particles are not a probable explanation for
the thin dust layers. We propose a natural explanation for the absense of
VSI in protoplanetary disks in Section \ref{sec-inhibiting-vsi-by-depletion},
and we finish with a discussion and conclusions.

\section{Stirring up of large dust aggregates by the VSI}\label{sec-stirring}
In this paper we wish to find out if the presence of the VSI in the outer
regions of a protoplanetary disk, such as the one around Oph 163131, would
inevitably lead to the big-grain dust layer observed with ALMA to be more
geometrically extended than observed. This would then be clear evidence that the
VSI does not operate in that disk.

The effect of the VSI on dust particles in the disk was studied by several
papers \citep[e.g.,][]{2015MNRAS.453L..78L, 2016A&A...594A..57S,
  2017ApJ...850..131F, 2022A&A...658A.156L}. The models we present in this
section are fundamentally different from those earlier papers. However,
we explore the parameters and compare the results to the observational
constraints.

\subsection{Conveyor-belt estimate of vertical mixing efficiency of the VSI}
\label{sec-conveyor-belt-estimate}
A simple estimate of the height above the midplane that a dust aggregate can be
lifted by the VSI would be the following. Assume that the VSI consists of
long-lived vertical upward and downward moving slabs of gas, acting as vertical
conveyor belts for the dust. As a dust aggregate gets dragged upward, the vertical
component of gravity increases linearly with $z$, leading to a vertical settling
of the aggregate with respect to the upward moving gas. The maximum height that
the dust aggregate can reach is the height $z$ at which the vertical settling
speed equals minus the vertical gas speed. The settling speed of a particle with
Stokes number $\mathrm{St}\ll 1$ at a height $z$ above the midplane is
\begin{equation}
v_{\mathrm{sett}} = -\mathrm{St}\,\Omega_K z \fullstop
\end{equation}
By setting $v_{\mathrm{sett}}+v_{z,\mathrm{VSI}}=0$, with $v_{z,\mathrm{VSI}}$ the
typical vertical gas velocities of the VSI, we obtain the maximum elevation
above the midplane that an aggregate can obtain:
\begin{equation}\label{eq-zmax-conveyor}
z_{\mathrm{max}} = \frac{h_p\,|v_{z,\mathrm{VSI}}|}{\mathrm{St}\,c_s} \comma
\end{equation}
where $h_p$ is the pressure scale height of the gas (see Appendix
\ref{app-fiducial-disk-model}) and $c_s$ is the isothermal sound speed of the
gas. For typical vertical velocities of the VSI of $|v_{z,\mathrm{VSI}}|\sim
0.1\,c_s$, a dust aggregate of $\mathrm{St}=0.1$ can be stirred up, according to
this simple estimate, to about one gas pressure scale height. If
Eq.~\ref{eq-zmax-conveyor} gives values larger than $h_p$, the estimate is
no longer accurate, and we limit it to $h_p$ for convenience.

In practice the mean elevation $\sqrt{\langle z^2\rangle}$ of the dust aggregate
will be smaller than this value, because the VSI motions are not stationary (see
Fig.~\ref{fig-vsi-vz-time}). But this estimate does explain why the VSI can stir
even large dust aggregates ($\mathrm{St}\simeq 1$) very far away from the midplane.

The conveyor-belt estimate can be compared to the more traditional
settling-mixing equilibrium (see Appendix \ref{sec-settling-mixing}).

\subsection{Particle motion model}
\label{sec-particle-motion-in-vsi-model}
A more accurate estimation of how dust aggregates are stirred up from the
midplane by the VSI is to compute their detailed motion within a hydrodynamic
model of the VSI. We employ the PLUTO code \citep{2007ApJS..170..228M} for
this. The setup of the disk follows the fiducial model of Appendix
\ref{app-fiducial-disk-model}.

We assume the disk to be perfectly locally isothermal and inviscid,
\revisedd{which we expect to maximize the VSI activity}. Given that
the VSI establishes itself primarily in the radial and vertical coordinates, we
model it in 2D using spherical coordinates $r$ and $\theta$. The radial
coordinate $r$ has 882 grid points logarithmically spaced between $0.2\,r_0$ and
$5\,r_0$, where $r_0=100\,\mathrm{au}$ is the reference radius. The vertical
coordinate $\theta$ (where $\theta=\pi/2$ is the equatorial plane) has 160 grid
points linearly spaced between \revised{$\pi/2-0.3$ and $\pi/2+0.3$}, which corresponds to
20 cells per scale height at $r=r_0$. This is enough to resolve the large-scale
structure of the VSI \citep{2020MNRAS.499.1841M}. At $r=r_0$ the range in
$\theta$ corresponds to $\pm\,4\,h_p$, dropping to $\pm\,2.7\,h_p$ at
$r=5\,r_0$.  The temperature is fixed in time, and depends on radius $r$ as
\revised{$T\propto r^{q}$ with $q=-1/2$}. It is chosen such that at $r=r_0$ the
gas pressure scale height is $h_p(r_0)=0.0732\,r_0$.

\begin{figure*}
  \centerline{\includegraphics[width=1\textwidth]{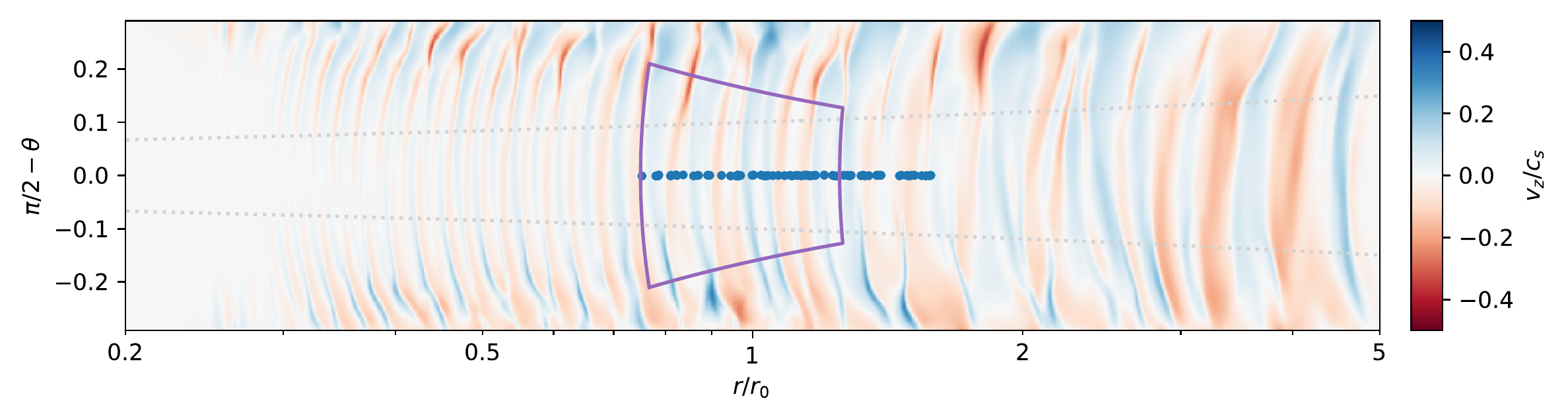}}
  \caption{\label{fig-fiducial-model}Vertical gas velocity $v_z$ in the disk
    model at time $t=300\,P_{\mathrm{orb}}(r_0)$, where $P_{\mathrm{orb}}(r_0)$
    is the orbital period at $r=r_0$. The coordinates are the natural spherical
    coordinates of the numerical hydrodynamic model: On the horizontal axis the
    natural logarithm of the spherical radius $r$ in units of $r_0$. On the
    vertical axis the polar angle $\pi/2-\theta$. Blue is upward and red is
    downward. The gray dotted lines show the gas pressure scale height. The blue
    dots are the initial locations of the particles, where only every 25th of
    the 2000 particles is shown. The purple box represents the zoom-in view
    shown in Fig.~\ref{fig-vsi-stirring-various-St}.}
\end{figure*}

\begin{figure*}
  \centerline{\includegraphics[width=1\textwidth]{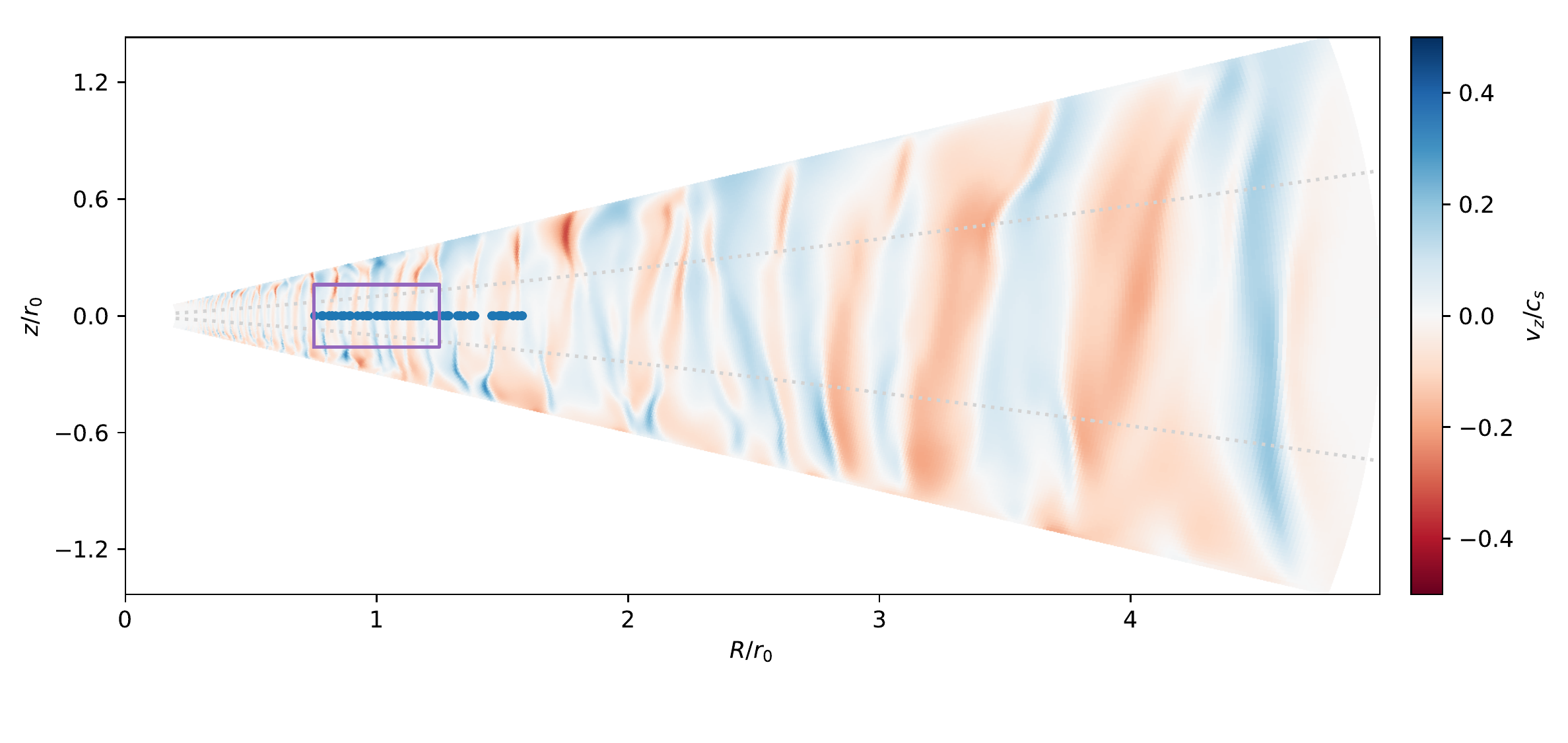}}
  \caption{\label{fig-fiducial-model-natural}Same as Fig.~\ref{fig-fiducial-model}, but
  now in natural (linear cylindrical) coordinates.}
\end{figure*}

\begin{figure}
  \centerline{\includegraphics[width=0.5\textwidth]{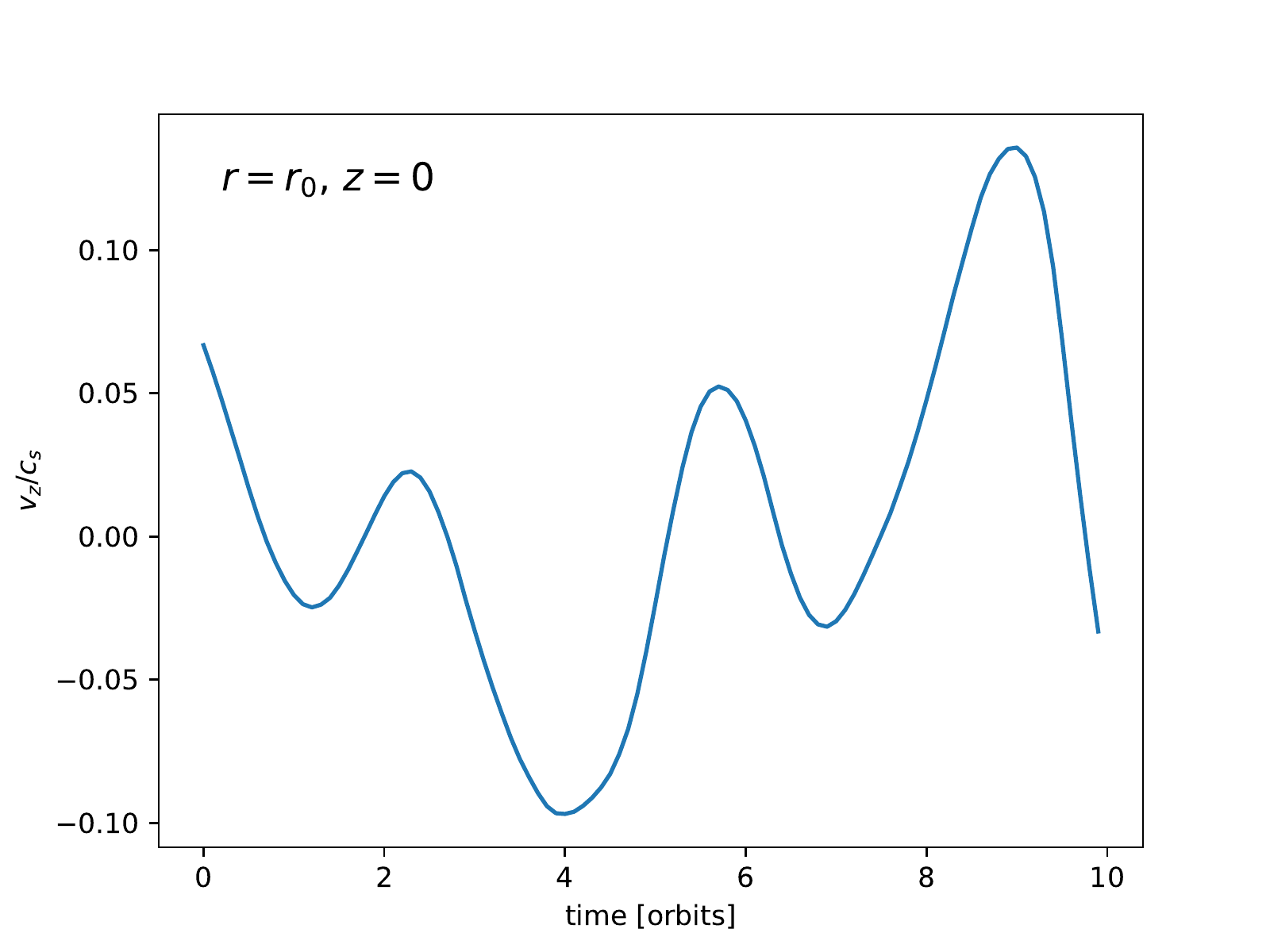}}
  \caption{\label{fig-vsi-vz-time}For the fiducial model shown in
    Figs.~\ref{fig-fiducial-model} and \ref{fig-fiducial-model-natural}, the
    vertical gas velocity $v_z$ at $r=r_0$ and $z=0$ in units of the local isothermal
    sound speed as a function of time in units of orbits after the 300th orbit.}
\end{figure}

We compute the gas dynamics without accounting for dust dynamics. Once the VSI
is fully developed, after about 300 orbits at $r=r_0$, we extract 200 time snapshots,
0.1 orbits apart in time. The vertical gas motions in the first of these
snapshots is shown in Fig.~\ref{fig-fiducial-model} for the entire radial and
vertical range of the model. The same is again plotted in
Fig.~\ref{fig-fiducial-model-natural} in natural (linear cylindrical)
coordinates, which gives a better view of the proportions. In Figure
\ref{fig-vsi-vz-time} the vertical gas velocity at the location $r=r_0$ and $z=0$
is shown as a function of time, to show how the VSI motions oscillate
with a period of a few local orbits.

For these 20 orbits we now follow, as a post-processing step, the motion of
$N=2000$ large dust particles that have been randomly placed between $0.75\,r_0$
and $1.6\,r_0$ radially, and between $-0.001\,r_0$ and $+0.001\,r_0$
vertically. The particle velocities are initialized as being equal to the local
Kepler velocity. The particles all have the same $\mathrm{St}_0$, meaning that
if they were placed at $r=r_0$ and $z=0$, they would have Stokes number
$\mathrm{St}=\mathrm{St}_0$. At each time step, for each particle, we recompute
$\mathrm{St}$ based on the local conditions, consistent with keeping the grain
size constant. The equations of motion of the particles include the force of
gravity as well as the friction with the gas. \revised{We implement the
  numerical integration of these equations in a Python program.} The
particles do not have dynamical feedback onto the gas, allowing the gas
hydrodynamics to be precomputed, and the dust particle dynamics to be computed
in post-processing mode.

\revised{Given that the particles, in spite their comparatively large size (of
  the order of $\sim$millimeter), are much smaller than the mean free path of
  the gas molecules, the friction force is the simple Epstein drag law. At each time
  step the local gas temperature, density and velocity are linearly interpolated
  in time and space from the precalculated 200 snapshots from the hydrodynamic
  simulation, using the RegularGridInterpolator function of the SciPy library.}

\begin{figure*}
  \centerline{\includegraphics[width=1\textwidth]{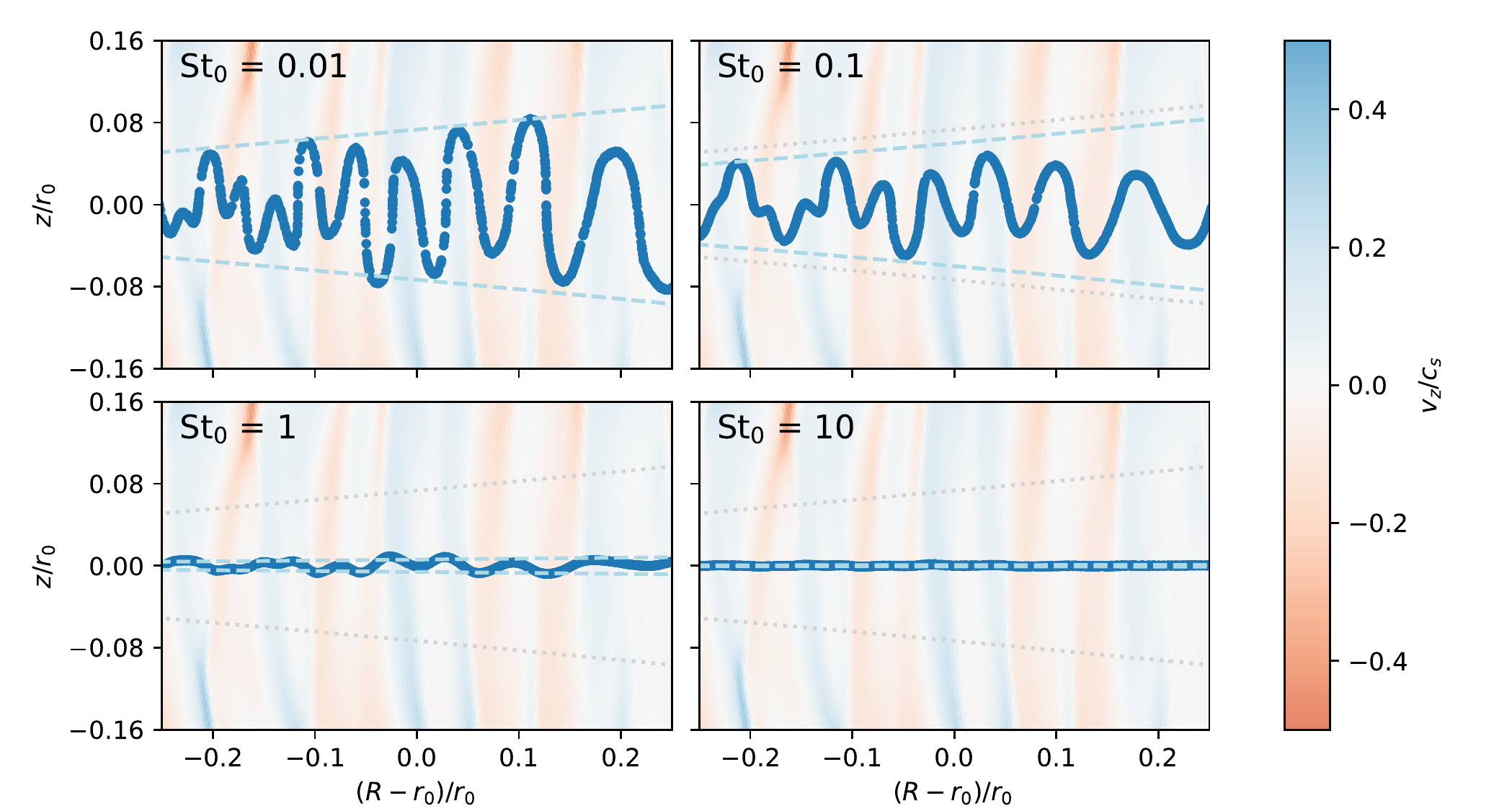}}
  \caption{\label{fig-vsi-stirring-various-St}Snapshots of the location of the
    particles (blue dots) 2.5 orbits after they were inserted at the midplane into the
    fully developed VSI hydrodynamic model, for four values of
    $\mathrm{St}_0$. The coordinates are the cylindrical radius $R$ in units of,
    and relative to, $r_0$, and the cylindrical vertical height $z$ above the
    midplane in units of $r_0$. The background image shows the vertical gas
    velocity $v_z$, where blue is upward and red is downward, in the same
    color scale as in Fig.~\ref{fig-fiducial-model}. The gray dotted
    lines show the gas pressure scale height for comparison. The lightblue dashed
    lines are the conveyor-belt estimate of the maximum vertical height of the dust
    particles, Eq.~(\ref{eq-zmax-conveyor}) with an upper cap at $z=h_p$ (which is
    why the dashed and dotted lines overlap for $\mathrm{St}_0=0.01$).}
\end{figure*}

In Fig.~\ref{fig-vsi-stirring-various-St} the results of the model are shown for
$\mathrm{St}_0=0.01$, $\mathrm{St}_0=0.1$, $\mathrm{St}_0=1$, and
$\mathrm{St}_0=10$, after 2.5 orbits at radius $r_0$. This short time is enough
to achieve approximately the typical heights above and below the midplane that the
particles acquire, and this does not change much in time after that. For
$\mathrm{St}\gtrsim 1$, the conveyor-belt estimate of the height
$z_{\mathrm{max}}$ above the midplane that the particles are stirred
(Eq.~\ref{eq-zmax-conveyor}) appears to be a reasonably good estimate, as can be
seen by comparing the vertical locations of the particles with the light-blue dashed lines in the figure. For
$\mathrm{St}=0.1$ it is not bad, but for smaller Stokes number the conveyor-belt
esimate strongly overestimates the stirring-up height of the particles. This is
because the VSI motions oscillate with a period of a few local Kepler orbits
(see Fig.~\ref{fig-vsi-vz-time}), which means that before the particles have
arrived at their maximum possible elevation, they will be transported back down
by the gas.

It is evident that the $\mathrm{St}_0=0.01$ particles are stirred up to \revised{one}
gas pressure scale height. This is entirely due to the VSI, and not
due to any $\alpha$-diffusion, which is not included in the model. For
$\mathrm{St}_0=0.1$, which is typically the highest Stokes number expected in
dust coagulation models in the outer disk regions \citep{2021A&A...647A..15D},
the dust particles are still stirred up to a substantial fraction of the gas
pressure scale height. Even for $\mathrm{St}_0=1$ the particles get up to a
height $z/R\sim 0.005$, which is about 7\% of the gas pressure scale
height, which is marginally consistent with the upper limit obtained for
Oph 163131. If we go to $\mathrm{St}_0=10$, the particles remain close to the
midplane, producing a thin layer well within the vertical geometric thinness
of the observed dust layer of Oph 163131. However, as is shown in Section
\ref{section-case-against-St-big}, it is unlikely that the particles in
this dust layer have $\mathrm{St}_0\gtrsim 1$.

\begin{figure*}
  \centerline{\includegraphics[width=0.5\textwidth]{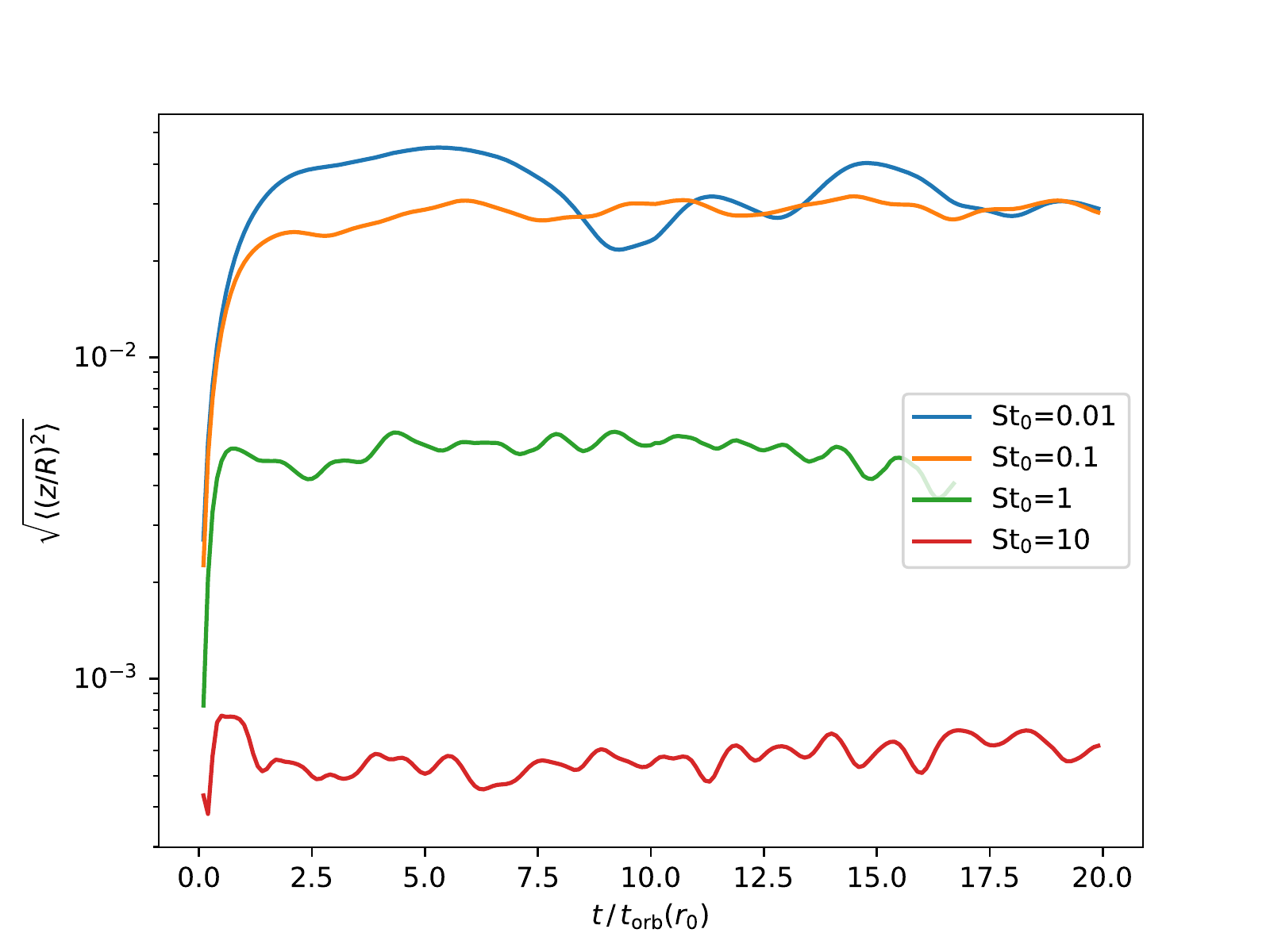}
  \includegraphics[width=0.5\textwidth]{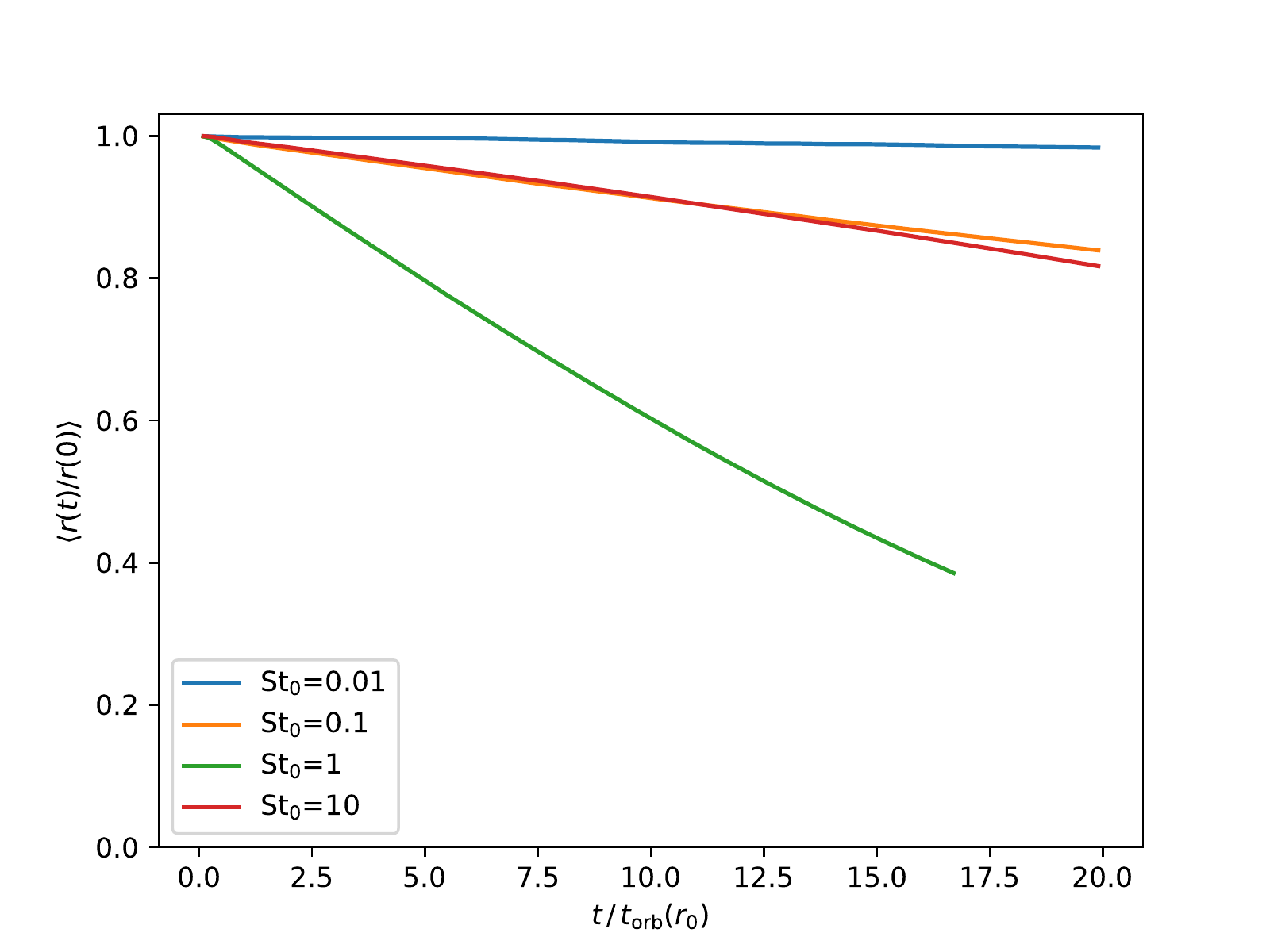}}
  \caption{\label{fig-vsi-stirring-mean-values-afo-time}Mean values of the
    particles as a function of time since their insertion into the hydrodynamics
    model. Left: The root-mean-square of the $z/r$ ratio of the particles. Right: the
    mean of the radius to initial radius ratio ($r(t)/r(t=0)$), as a measure of how
    strongly the particles have radially drifted inward since the time of insertion.
    For the $\mathrm{St}_0=1$ case the computation was halted after 17 orbits, when
    the first particles left the inner orbit of the model.}
\end{figure*}

To see this more quantitatively, we compute the root-mean-square of the $z/r$
ratio of the particles $\sqrt{\langle(z/r)^2\rangle}$ as a function of time
since insertion. This is a measure of the vertical extent of the dust layer. The
results are shown in Fig.~\ref{fig-vsi-stirring-mean-values-afo-time}-left. As
can be seen, for $\mathrm{St}_0\ge 0.1$ the particles quickly reach a
steady-state vertical extent, which is smaller for larger values of
$\mathrm{St}_0$. 

For $\mathrm{St}_0=1$ the radial drift of the particles becomes so fast
that after 17 orbits the first particles leave the grid of the model at the
inner edge. The calculation is then halted.

In general it should be noted that the rapid radial drift of large dust
particles is a long-standing problem in the interpretation of millimeter wave
observations of protoplanetary disks \citep{2009A&A...503L...5B}. One
explanation could be that the gas disks are so massive, that even millimeter
particles in the outer regions of protoplanetary disks do not drift at excessive
speeds \citep{2017ApJ...840...93P}. Another explanation is that radial drift of
these particles is inhibited by dust trapping in one or more local pressure
maxima \citep{2012A&A...538A.114P}. This would imply that the dust we observe
with ALMA in the outer regions of protoplanetary disks ($r\gtrsim
20\,\mathrm{au}$) is either trapped in vortices or in rings. Both features are
indeed observed with ALMA in numerous disks \citep[e.g.,][]{2013Sci...340.1199V,
  2018ApJ...860..124D, 2015ApJ...808L...3A, 2018ApJ...869L..42H} and thus lend
support to this picture. This means that the very flat dust midplane layers, if
they consist of a series of concentric rings, could very well be made up of
large dust particles, without experiencing the strong radial drift that the
particles in our model undergo.

In principle this means that our models should be repeated for the case of disks
with radial pressure bumps. However, since the origin of these pressure bumps
is not yet clear, this would introduce a series of new and unconstrained model
parameters. Also, a variety of additional phenomena could occur in these traps
\citep[e.g.,][]{2021AJ....161...96C, 2022A&A...658A.156L}. So for this paper
we limit ourselves to disks without pressure traps.

\subsection{Dynamics of dust modeled as a fluid}
\label{sec-dust-fluid-motion-in-vsi-model}
The dust motion can also be modeled directly within the hydrodynamics model.
For this we employ the Fargo3D code\footnote{\url{http://fargo.in2p3.fr}}
\citep{2016ApJS..223...11B}, which has dust dynamics built in
\citep{2020RNAAS...4..198K}. \revisedd{Like before, we assume a locally
  isothermal equation of state, maximizing the VSI activity.} The dust is
treated as a pressureless fluid, which feels friction with the gas.  \revised{In
  the standard setup of Fargo3D, the gas feels the opposite force from the
  dust. However, to make the comparison with the results of Section
  \ref{sec-particle-motion-in-vsi-model}, we switch this feedback off. It is
  known that for high metallicity $Z$ the VSI can be hampered simply by the mass
  of the dust \citep{2020A&A...635A.190S,2022A&A...658A.156L}, which would be
  one possible explanation for the razor-thin dust disks seen in ALMA. But in
  this section we assume that this effect is not taking place.} The background
viscosity is set to $\alpha=10^{-6}$.

\begin{figure*}
  \centerline{\includegraphics[width=0.8\textwidth]{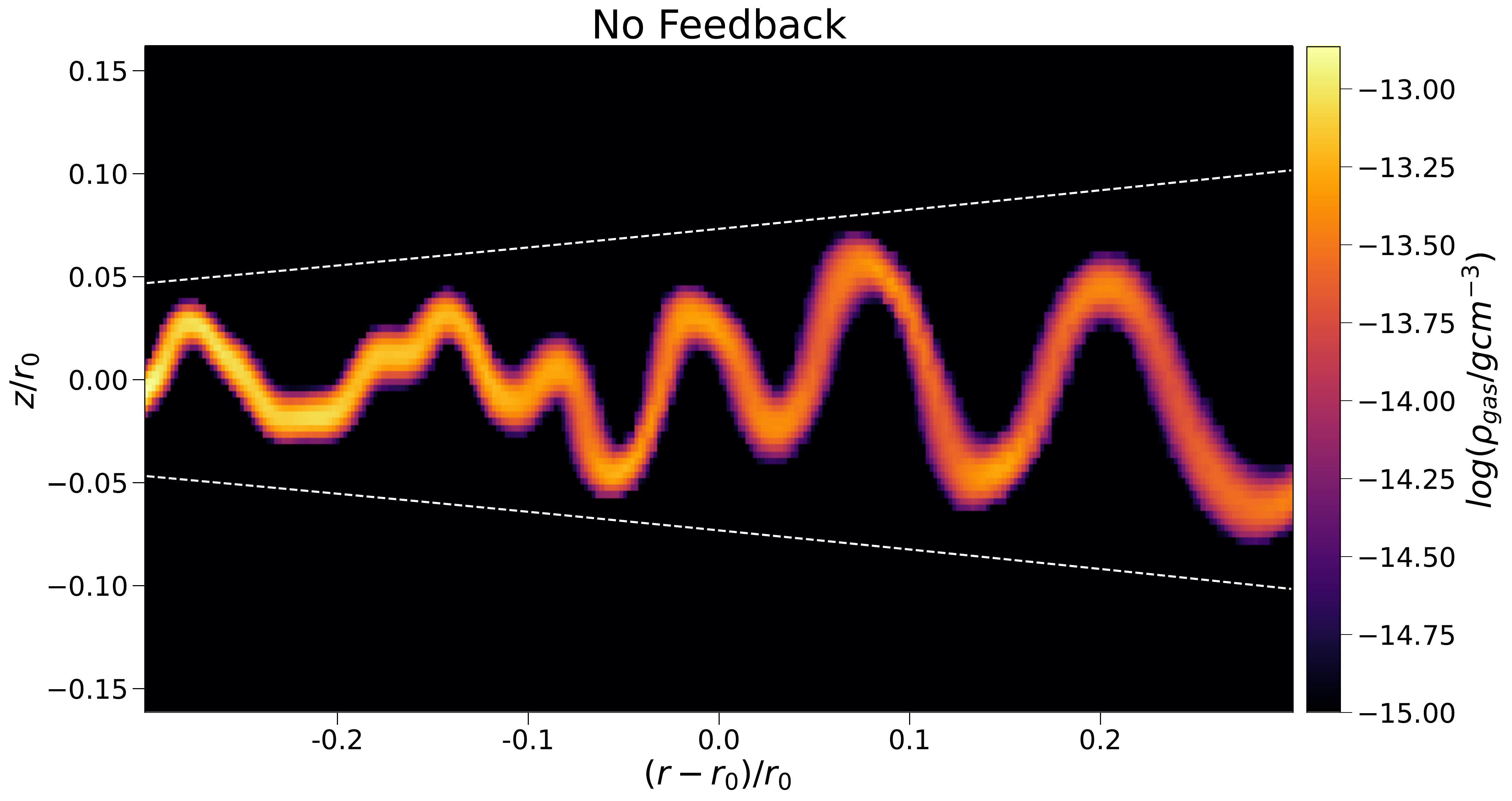}}
  \caption{\label{fig-real-dustmodel-St01}Spatial distribution of $\mathrm{St}_0=0.1$
    dust in the hydrodynamic model in which both gas and dust are dynamically
    modeled as a fluid (Section \ref{sec-dust-fluid-motion-in-vsi-model}). The white dashed lines
  mark one gas pressure scale height above/below the midplane.}
\end{figure*}

The results for the case of $\mathrm{St}_0=0.1$ are shown in
Fig.~\ref{fig-real-dustmodel-St01}. The dust has been allowed to settle from the
very beginning of the simulation over the entire modeling time. Throughout this
time frame, the pattern of the dust remains corrugated. It is very comparable to
the results of Section \ref{sec-particle-motion-in-vsi-model}. The main
difference is that in the dust-fluid approach the vertical width of the
corrugated dust ``layer'' is thicker than in the particle approach of Section
\ref{sec-particle-motion-in-vsi-model}. This is due to numerical diffussivity.

\begin{figure*}
  \centerline{\includegraphics[width=0.47\textwidth]{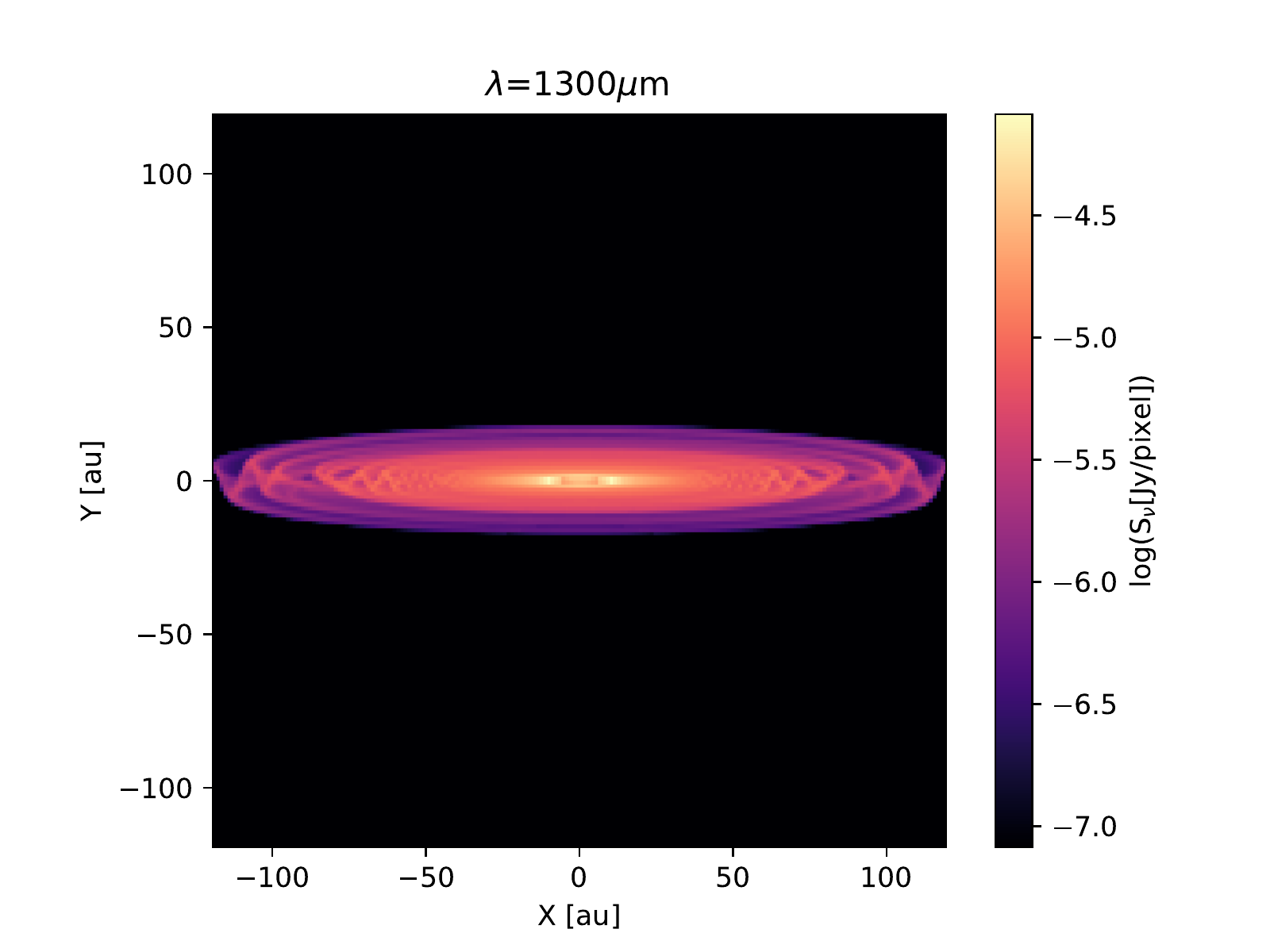}
    \includegraphics[width=0.47\textwidth]{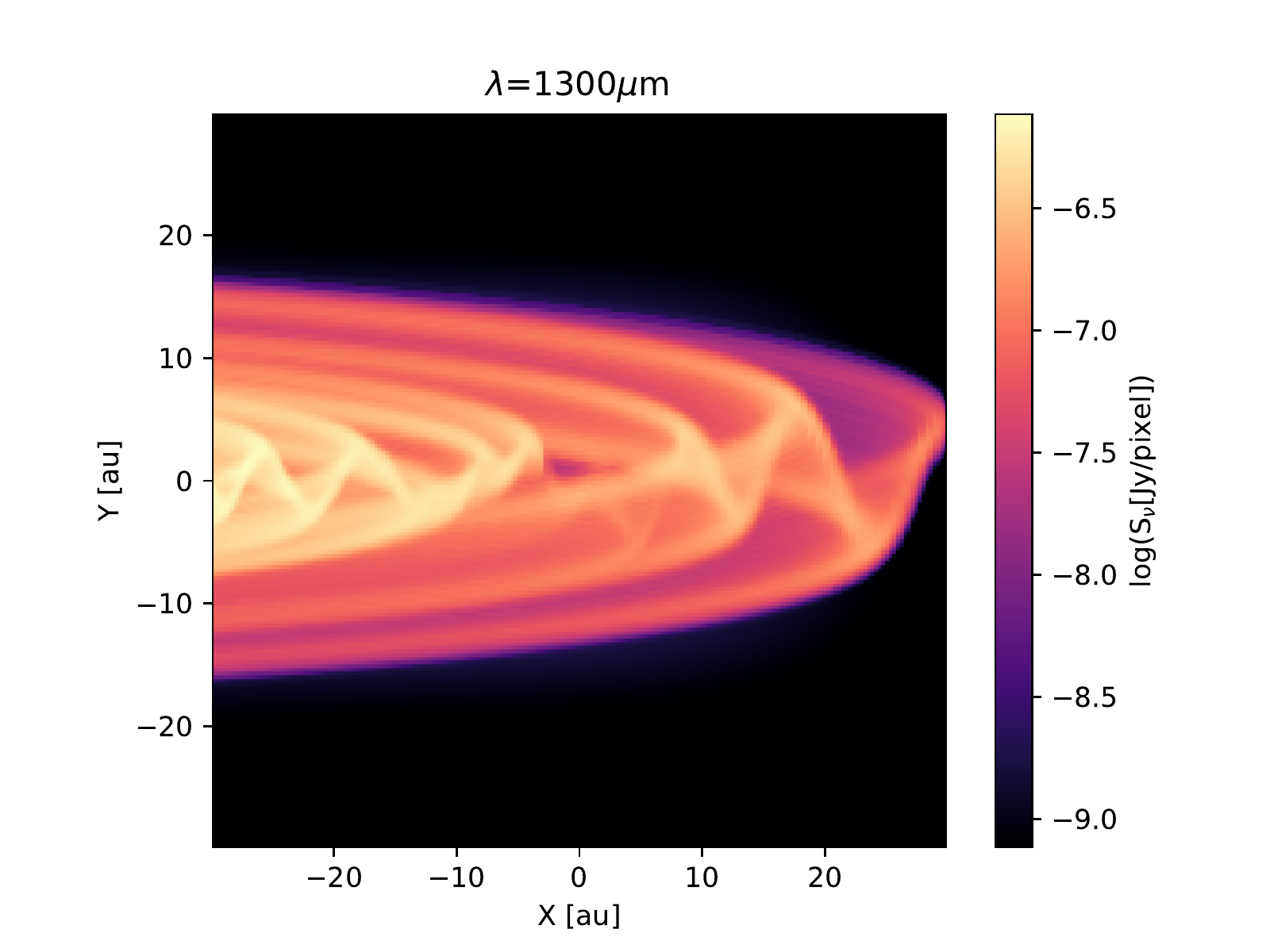}
  }
  \caption{\label{fig-vsi-alma}\revised{Synthetic image of the model of Section
    \ref{sec-dust-fluid-motion-in-vsi-model} at $\lambda=1.3$ mm at an
    inclination of 84$^\circ$. Right panel is a zoom in (color-rescaled) of the
    left panel. The gap was added artificially to demonstrate how the VSI
    affects the observability of a gap. The images are idealized: no ALMA beam
    convolution is applied.}}
\end{figure*}

\revised{Next, we put the result of this model into the
  RADMC-3D\footnote{\url{https://www.ita.uni-heidelberg.de/~dullemond/software/radmc-3d/}}
  radiative transfer code and compute the images at an inclination of
  84$^\circ$, at a wavelength of $\lambda=1300\,\mu$m. The big grains were
  assumed to have a radius of 100 $\mu$m, and we used the corresponding opacity
  for them (see Appendix \ref{app-dust-opacity-model}). The results are shown in
  Fig.~\ref{fig-vsi-alma}. The corrugated geometry of the dust ``layer'' is
  clearly seen. To stress the effect this has on identifying any potential
  radial gaps in the dust layer, we artificially added a gap between 87 au and
  98 au and a slight reduction of the density between 3.5 au and 60 au according to the model of
  \citet{2022arXiv220400640V} for Oph 163131. This was done a-posteriori: the
  big-grain dust density from the hydrodynamic model of Fargo3D was multiplied
  by a radial function that reduces the density by a factor of 0.1 between 87-98
  au and by 0.5 inward of 60 au. After that, it was inserted into the RADMC-3D
  code. As seen in Fig.~\ref{fig-vsi-alma}, these features are not recognizable
  due to the strong vertical waves.  These images are not convolved with the
  ALMA beam, as they merely serve as an illustration. For the case of Oph
  163131, \citet{2022arXiv220400640V} show that the spatial resolution of ALMA
  easily suffices to rule out that the dust layer is as strongly corrugated as
  in Fig.~\ref{fig-vsi-alma}.}

\begin{figure*}
  \centerline{\includegraphics[width=0.99\textwidth]{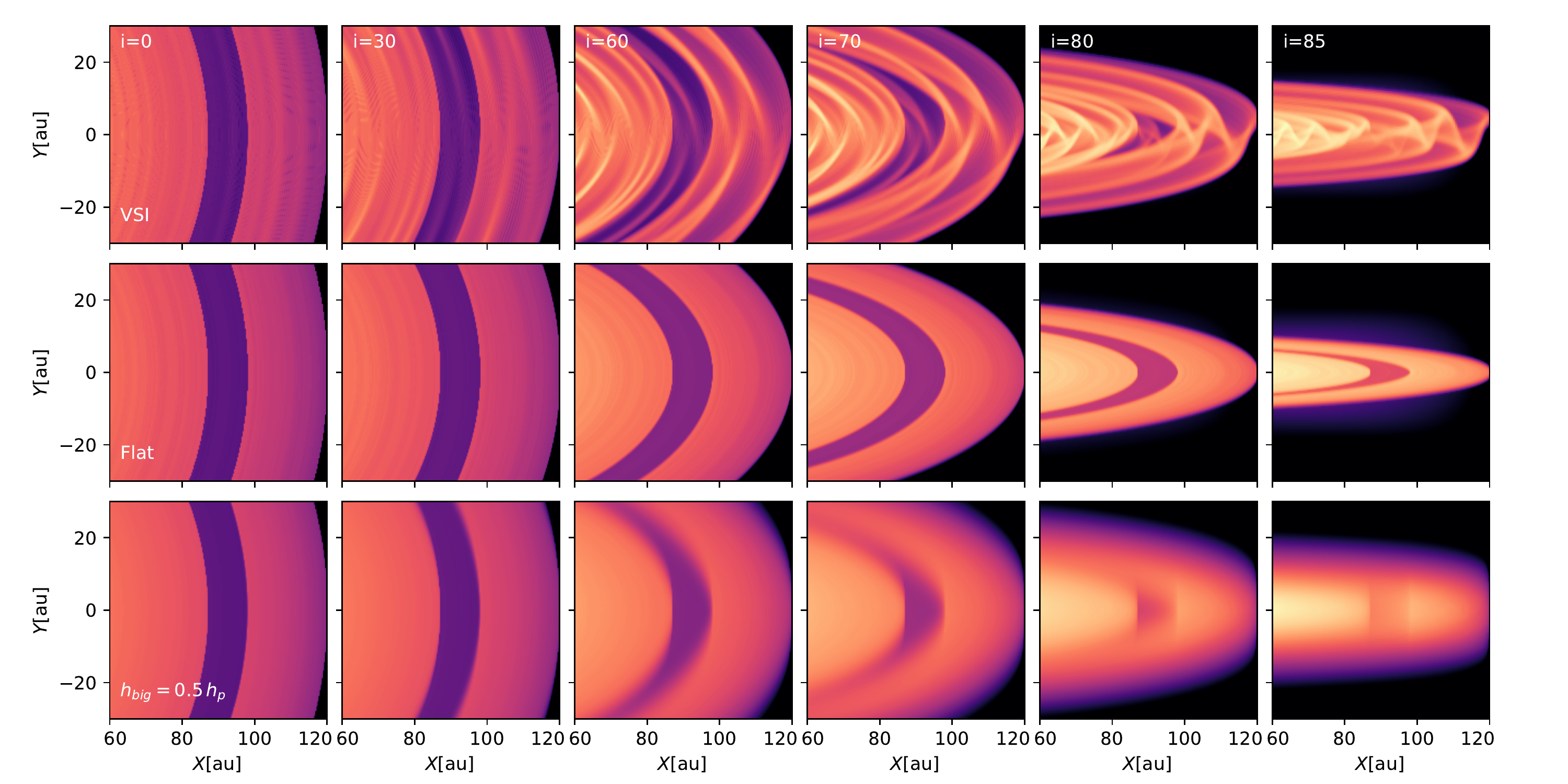}}
  \caption{\label{fig-vsi-and-flat-alma-incl}\revised{Comparison of the VSI model of
    Section \ref{sec-dust-fluid-motion-in-vsi-model} (top), a model with a flat
    dust layer (middle), and a model with a vertically extended dust layer with
    half the vertical thickness of the gas (bottom). Left-to-right: different
    inclinations. The synthetic observations are computed at $\lambda=1.3$ mm,
    and the images are centered 90 au offset from the star. No beam convolution
    is applied.}}
\end{figure*}

\revised{To highlight the difference between the VSI model and an equivalent
  model with a flat midplane layer, we show in
  Fig.~\ref{fig-vsi-and-flat-alma-incl} the comparison between these cases, for
  several inclinations. Again, no beam convolution is applied. Although
  the disk around Oph 163131 is used as a basis for these models, they are not
  meant to directly fit Oph 163131, but instead to illustrate the
  typical protoplanetary disk case (hence the different inclinations shown). It
  is clearly seen that at high inclinations, the VSI models look very different
  from the flat models, at scales easily resolvable with ALMA for objects at
  typical distances of about 100 pc. Also shown is the case where the big dust
  grains are vertically smeared out in a Gaussian layer with a vertical
  thickness half that of the gas ($h_{\bigdust}=0.5\,h_p$). This mimicks the
  case when the vertical dust transport by the VSI would be treated as a
  vertical turbulent mixing instead of an actual advective transport. This
  case looks also substantially different from the VSI case. But it will depend on
  the distance of the object and the ALMA baselines whether they can be
  distinguished. The differences become less clear at lower inclinations,
  because the models only differ in vertical direction and are
  the same in radial direction.}

\revised{A caveat of these synthetic images of the VSI-stirred dust disk is that
  we have inserted only a single grain size. If we assume that the large grains
  follow a size distribution of a certain width, then the strong wiggles seen in
  the image get smeared out. The degree of smearing-out depends on the width of
  the size distribution. But it would not affect the conclusions of this paper.}

This simulation confirms the results of Section
\ref{sec-particle-motion-in-vsi-model} that even particles with a rather high
Stokes number of $\mathrm{St}_0=0.1$ are stirred up to a substantial fraction of
the gas pressure scale height, easily measureable with ALMA, and clearly in
contrast with for instance the ALMA observations of Oph 163131.

As in the models of Section \ref{sec-particle-motion-in-vsi-model}, the
corrugated pattern of the dust layer is not static. It follows the
time-dependent variations of the VSI velocity profile, where upward gas motions
turn into downward motions and vice versa over time scales of a few local
orbits. The model also confirms that the dust is not ``vertically mixed'' as in
the simple vertical mixing-settling model of \citet{1995Icar..114..237D},
\citet{2004A&A...421.1075D} and \citet{2009A&A...496..597F}. Instead, the
corrugated structure of the dust is maintained, and the vertical extent is
better described by the ``conveyor-belt model'' of Section
\ref{sec-conveyor-belt-estimate}.

\subsection{Conclusion of this section}
In this section we have shown that with a VSI operating in the disk, even
particles with a Stokes number close to unity get stirred up to high elevations
above the midplane, of the order of the gas pressure scale height. This is
in conflict with ALMA observations of several protoplanetary disks, most
strikingly the disk around Oph 163131 \citep{2022arXiv220400640V}.

However, for $\mathrm{St}_0\gg 1$ the midplane dust layer indeed becomes
very geometrically thin, even in a disk in which the VSI is operating. So
we need to rule out that these particles could have $\mathrm{St}_0\gg 1$,
which is the topic of Section \ref{section-case-against-St-big}.

Once we ruled it out, we have to investigate how the VSI could be suppressed in
the outer regions of protoplanetary disks. This is explored in Section
\ref{sec-inhibiting-vsi-by-depletion}.

\section{The case against the midplane dust layer consisting of $\mathrm{St}\gg 1$ particles}
\label{section-case-against-St-big}
As Section \ref{sec-stirring} showed, the geometric thinness of the midplane
dust aggretate layers in protoplanetary disks can most easily be explained by
particles that have $\mathrm{St}\gtrsim 1$, since they remain in a thin layer in
spite of a possible VSI operating in the background. However, the dust rings
seen in ALMA observations at $\lambda=1.3\,\mathrm{mm}$ tend to have optical
depths larger than about $0.3$ at that wavelengths \citep{2018ApJ...869L..46D}.
For Oph 163131 in particular, \citet{2022arXiv220400640V} find with their radiative
transfer modeling that the midplane dust layer is partially optically thick.

We demonstrate in this section that having both $\mathrm{St}\gtrsim 1$ and
\revised{$\tau_{1.3\,\mathrm{mm}}\gtrsim 0.3$} requires a vertically integrated dust-to-gas
ratio of \revised{at least $Z\gtrsim 0.08$, but likely $Z\gtrsim 0.16$.}
This value increases linearly with increasing
$\mathrm{St}$ and $\tau_{1.3\,\mathrm{mm}}$. 

\begin{figure}
  \centerline{\includegraphics[width=0.5\textwidth]{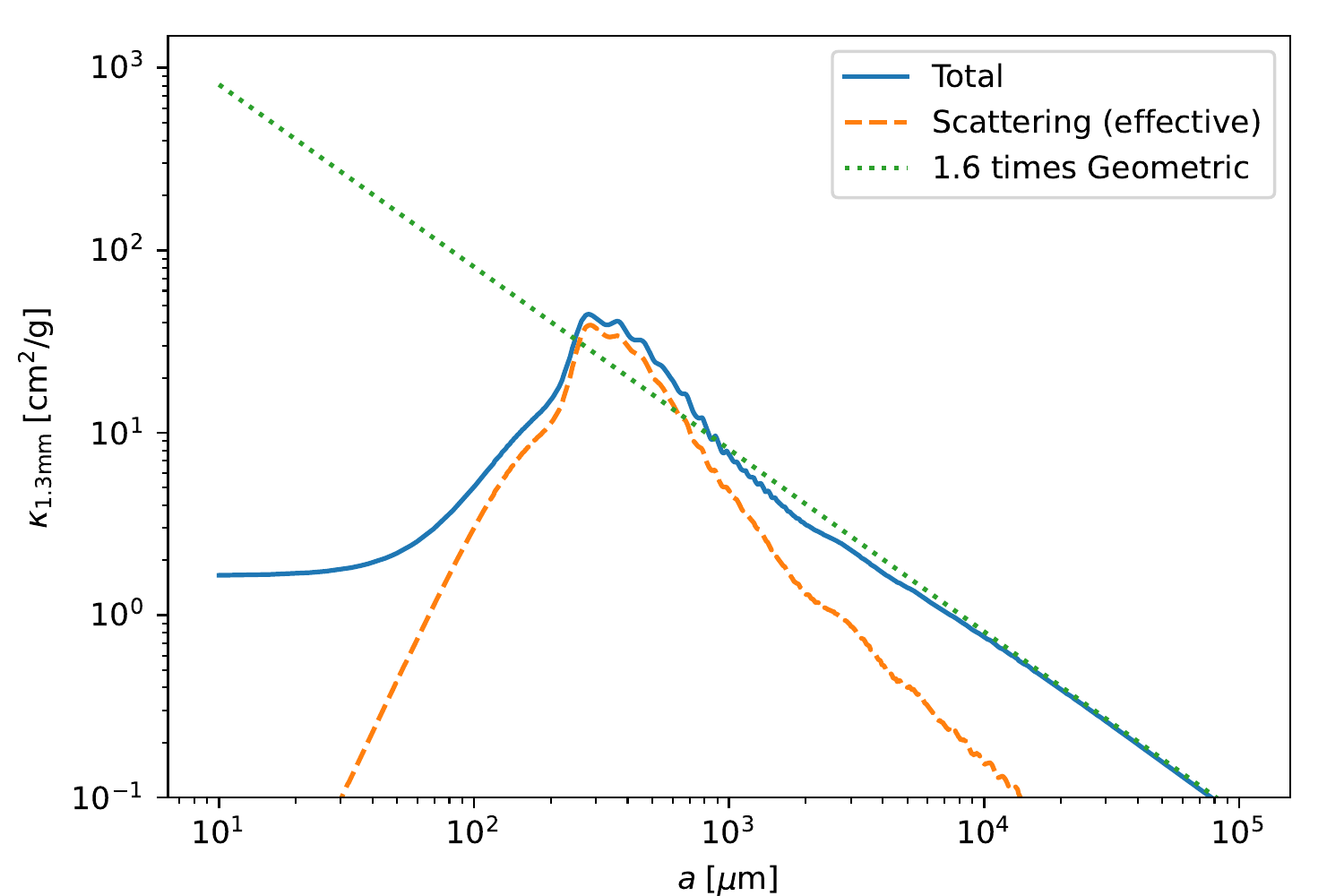}}
  \caption{\label{fig-kappa-mono}Dust opacity at $\lambda=1.3\,\mathrm{mm}$ as
    a function of grain size \revised{in units of cm$^2$ per gram of dust}.
    Solid line: total opacity
    $\kappa_{1.3\,\mathrm{mm}}^{\mathrm{tot}}$; dashed line: effective
    scattering opacity $\kappa_{1.3\,\mathrm{mm}}^{\mathrm{scat,eff}}$; dotted
    line: Eq.~(\ref{eq-kappa-as-kgeom}) for $\Qlambda=1.6$.}
\end{figure}

To arrive at this, we start with the dust opacity model described in
Appendix \ref{app-dust-opacity-model}. In Fig.~\ref{fig-kappa-mono} the
$\lambda=1.3\,\mathrm{mm}$ opacity as a function of grain size for this dust
model is shown. Clearly the opacity $\kappa_{1.3\,\mathrm{mm}}$ is a strong
function of the grain size $a$. It has a maximum value of \revised{44.6} cm$^2$/g at a
grain size of $a=0.28\,\mathrm{mm}$. For $a\rightarrow 0$ the asymptotic value
is $\kappa_{1.3\mathrm{mm}}\rightarrow$ \revised{1.65} cm$^2$/g. For $a\rightarrow \infty$
we can express $\kappa_\lambda$ in terms of the geometric opacity
$\kappa_{\mathrm{geom}}$ defined as the geometric cross section $\pi a^2$
divided by the grain mass:
\begin{equation}\label{eq-kappa-as-kgeom}
\kappa_\lambda = \Qlambda \kappa_{\mathrm{geom}} = \frac{3}{4}\frac{\Qlambda}{\rho_s a} \comma
\end{equation}
where $\rho_s$ is the mean material density of the dust aggregate, and $\Qlambda$
is the ratio of the opacity to the geometric
opacity \revised{\citep{vandehulstlightscattering}}.
For $a\gtrsim 1\,\mathrm{mm}$, Eq.~(\ref{eq-kappa-as-kgeom})
provides a good fit to the real opacity for a constant $\Qlambda=1.6$. We note,
however, that this equation can be used for all values of $a$, in which case
$\Qlambda$ will depend on $a$ and drop well below unity for $a\ll \lambda/2\pi$.

If we define the surface density of the big dust grains as $\Sigma_{\bigdust}$,
then the optical depth of the big dust grain
layer becomes
\begin{equation}\label{eq-tau-bd-from-sigma}
\tau_{\lambda,\bigdust} = \Sigma_{\bigdust}\kappa_\lambda =
  \frac{3}{4}\frac{\Sigma_{\bigdust}\Qlambda}{\rho_s a} \comma
\end{equation}
where $a$ is the radius of the big dust grains.

Next, let us compute, for the big dust grains, the Stokes number \St.  For the
outer regions of the protoplanetary disk we can assume that we are firmly in
the Epstein friction regime, so that we can write \citep{2010A&A...513A..79B}:
\begin{equation}\label{eq-stokes-from-a}
\St_{\bigdust} = \frac{\pi\rho_s a}{2\Sigma_{\mathrm{g}}} \comma
\end{equation}
where $\Sigma_{\mathrm{g}}$ is the surface density of the gas. We can combine
Eqs.~(\ref{eq-tau-bd-from-sigma}, \ref{eq-stokes-from-a}) and eliminate
$\rho_s a$, to obtain
\begin{equation}\label{eq-z-from-st-and-tau}
  Z_{\bigdust}\equiv
  \frac{\Sigma_{\bigdust}}{\Sigma_{\mathrm{g}}} =\frac{8}{3\pi}
  \frac{\St_{\bigdust}\tau_{\lambda,\bigdust}}{\Qlambda} \fullstop
\end{equation}
In the dust opacity model of Appendix \ref{app-dust-opacity-model}, for
$\lambda=1.3\,\mathrm{mm}$, the maximum value of $\Qlambda$ is \revised{3.0}, which is
only reached in a narrow range of grain radii (see Fig.~\ref{fig-kappa-mono}).
For most values of $a$, $\Qlambda\lesssim 1.6$. This means that if both
$\St_{\bigdust}\gtrsim 1$ and \revised{$\tau_{\lambda,\bigdust}\gtrsim 0.3$}, then
Eq.~(\ref{eq-z-from-st-and-tau}) shows that {\revised{$Z_{\bigdust}\gtrsim 0.08\cdots 0.16$}, i.e.,
the ``metallicity'' must be extremely high.

The question is then: is this a realistic scenario? Can the geometrically thin,
but optically \revised{marginally} thick \revised{($\tau\gtrsim 0.3$)} dust
rings seen in many protoplanetary disks (and most strikingly seen in Oph 163131)
be rings of dust particles with $\St_{\bigdust}\gtrsim 1$ and
\revised{$Z_{\bigdust}\gtrsim 0.16$}, or even $\St_{\bigdust}\gg 1$ and
$Z_{\bigdust}\gg 1$, with a dynamics similar to the rings of Saturn? This
scenario is completely different from the standard picture of dust dynamics in
protoplanetary disks, which assume $\St_{\bigdust}\ll 1$ and $Z_{\bigdust}\ll
1$.

Although we cannot rule it out, we consider \revised{this scenario unlikely.
  The conditions derived in this section are the minimal conditions
  required. To be more comfortably within the limits, one would need
  $Z_{\bigdust}/\St_{\bigdust}\gg 0.16$, leading, for $\St_{\bigdust}\gtrsim 1$,
  to very large values of $Z_{\bigdust}$.}


\revised{Using this constraint, we are then forced to consider mechanisms
  quenching the VSI entirely in order to understand the geometrical thinness of
  the dust rings. One way would be to load so much mass worth of dust in this
  midplane layer, that the gas is no longer able to lift it up from the
  midplane. As was shown by \citet{2019MNRAS.485.5221L},
  \citet{2020A&A...635A.190S} and \citet{2022A&A...658A.156L}, the VSI is
  suppressed if the vertically integrated dust-to-gas ratio (``metallicity'') of
  the big grains exceeds about $Z\gtrsim 0.02\cdots 0.05$. Another way is to
  increase the cooling time scale, which is a natural consequence of grain
  growth \citep{2021ApJ...914..132F}.}

\section{Inhibiting the VSI through depletion of small dust grains by grain growth}
\label{sec-inhibiting-vsi-by-depletion}
The VSI operates in disks which have, to good approximation, a locally
isothermal equation of state. That is, at any given position $(\rcyl,z,\phi)$ in
the disk the temperature of the gas $T_g(\rcyl,z,\phi)$ is fixed and does not
vary in time. The justification for this is that in the outer regions of
protoplanetary disks, the thermal household is determined by \revised{a balance
  between} irradiation from the central star \revised{and thermal radiative
  cooling by the dust}. \revised{The radiative cooling time scale
  $t^{\mathrm{rad}}_{\mathrm{cool}}$
  for any perturbation of this equilibrium is} short compared to the orbital
time scale.

However, the cooling time is not completely negligibly small compared to the
orbital time scale. As we shall show, only a moderate amount of dust coagulation
is enough to increase the cooling time \revised{(or more accurately, the
relaxation time; see Appendix \ref{app-thermal-relaxation-time-thin})}
beyond the limit where the VSI is stopped.

It was shown by \citet{2015ApJ...811...17L} that if the \revised{thermal
  relaxation} is not fast enough, \revised{the vertical entropy gradient}
acts as a strongly stabilizing force against the VSI. They derive the following
upper limit on the radiative \revised{relaxation} time $t_{\mathrm{relax}}$:
\begin{equation}\label{eq-vsi-upper-limit-tcool}
t_{\mathrm{relax}}<\frac{|q|}{\gamma-1}\frac{1}{\Omega_K}\frac{\hpgas}{r} \comma
\end{equation}
where $q$ is the powerlaw index of the midplane temperature profile of the disk
$\tmid\propto r^{q}$, and $\gamma$ is the usual adiabatic index of the gas,
which for the outer disk regions is $\gamma=5/3$ because the rotational and
vibrational modes of H$_2$ are not excited at those temperatures. For our
fiducial disk model (Appendix \ref{app-fiducial-disk-model}) we have $q=-1/2$,
and at $r=r_0=100\,\mathrm{au}$ we have $\hpgas/r=0.0732$. So we obtain
$t_{\mathrm{relax}}<0.055/\Omega_K$ as the upper limit on the
\revised{thermal relaxation} time for VSI to be operational.

Where in the protoplanetary disk this condition is met, and where not, was,
among other things, explored by \citet{2019ApJ...871..150P}. They found that the
VSI is typically operational for radii $\rcyl\gtrsim 10\,\mathrm{au}$, which are
the regions of protoplanetary disks that have been resolved with ALMA, and where
these geometrically thin dust layers are detected.

\revised{\citep{2021ApJ...914..132F} explore how dust evolution can change this,
  and they found that the coagulation of dust grains can increase the relaxation
  time scale and act against the VSI. A similar conclusion for \revisedd{the
    Zombie Vortex Instability} was found by \citet{2018ApJ...869..127B}.}

In this section we revisit this, and estimate $t_{\mathrm{relax}}$ in a
simplified, \revised{yet robust way, including realistic dust opacities and} the
effect of dust depletion due to coagulation. We mimick the effect of
coagulation by a simple conversion factor $\coag\in [0,1]$ that says that a
fraction $\coag$ of the small grains has been converted into big grains that are
not participating in the radiative cooling of the gas (these are probably the
grains we observe with ALMA), while only a fraction $(1-\coag)$ of the small
grains remain to radiatively cool the gas. \revised{In essence, we make the
  simplifying assumption that the dust consists of only two components: small
  submicron dust grains that are well-mixed with the gas, and are solely
  responsible for the radiative cooling, and big millimeter-size grains that
  tend to settle to the midplane unless they are stirred up by the gas.}

\subsection{Gas cooling via small dust grain emission}
\label{sec-gas-cooling-dust-radiation}
In the outer regions of a protoplanetary disk, the gas near the disk midplane is
cold: $\tmid\lesssim 70\,\mathrm{K}$. This means that the gas has very few
emission lines, and no continuum, by which it can radiatively cool: typically
only the rotational transitions of CO and its isotopologs, and maybe a few
more complex molecules. Effectively this means that the gas is unable to
radiatively cool by itself. It can only cool by transmitting its thermal energy
to the available dust grains in the gas, which then can radiate away this
energy.

In the midplane regions of the disk, the thermal coupling of gas and dust
through collisions of gas molecules with the dust particles, is relatively
efficient, though not perfect. The gas-dust thermal coupling time scale is estimated in Appendix
\ref{app-thermal-coupling-dust-gas}, but \revised{first} we assume that the
gas and dust thermally equilibrate fast enough that we can set $T_g=T_{\smalldust}$,
i.e., the gas temperature equals the small-grain dust temperature.

We ignore the effect of the large-dust-aggregates midplane layer, and focus only
on the gas and the small dust grains floating in the gas. We
assume that these small dust grains are well-mixed with the gas in vertical
direction, so that the dust-to-gas ratio for these small dust grains is
vertically constant.

Under these conditions, the fastest cooling happens in the optically thin
regime, because the dust opacity is independent of the dust density (the amount
of dust per unit volume of the disk).

The rate of thermal emission of the small dust grains per unit volume of the disk
is:
\begin{equation}\label{eq-q-cool}
q_{\mathrm{cool,\smalldust}} = 4\pi \rho_d \int_0^\infty \kappa^{\mathrm{abs}}_{\nu,\smalldust} B_\nu(T_{\smalldust})d\nu \comma
\end{equation}
where $\rho_d$ is the volume mass density of the small dust grains,
$\kappa^{\mathrm{abs}}_{\nu,\smalldust}$ is their absorption opacity as a function of
frequency $\nu$, and $B_\nu(T_{\smalldust})$ is the Planck function at the dust temperature
$T_{\smalldust}$. It is convenient to express this in terms of the Planck mean opacity
$\kappa_P(T_{\smalldust})$ defined as
\begin{equation}\label{eq-define-planck-mean}
\begin{split}
  \kappa_P(T_{\smalldust}) &= \frac{\int_0^\infty \kappa^{\mathrm{abs}}_{\nu,\smalldust} B_\nu(T_{\smalldust})d\nu}{\int_0^\infty B_\nu(T_{\smalldust})d\nu}\\
  &= \frac{\pi}{\sigma_{\mathrm{SB}}T_{\smalldust}^4}\int_0^\infty \kappa^{\mathrm{abs}}_{\nu,\smalldust} B_\nu(T_{\smalldust})d\nu \comma
\end{split}
\end{equation}
with $\sigma_{\mathrm{SB}}$ the Stefan-Boltzmann constant. We can then
express $q_{\mathrm{cool,\smalldust}}$ as
\begin{equation}\label{eq-q-cool-kappl}
q_{\mathrm{cool,\smalldust}} = 4\rho_{\smalldust} \kappa_P(T_{\smalldust})\,\sigma_{\mathrm{SB}}\,T_{\smalldust}^4 \fullstop
\end{equation}
In Appendix \ref{app-planck-mean-opacity} we give a convenient approximate
expression for $\kappa_P(T_{\smalldust})$. 

The thermal energy in the dust per unit volume of the disk is:
\begin{equation}\label{eq-e-th-smalldust}
e_{\mathrm{th,\smalldust}} = c_{V,\smalldust}\,\rho_{\smalldust}\,\,T_{\smalldust} \comma
\end{equation}
with $c_{V,\smalldust}$ the \revised{specific} thermal heat capacity of the dust, 
$c_{V,\smalldust}\lesssim 10^7\,\mathrm{erg}\,\mathrm{g}^{-1}\,\mathrm{K}^{-1}$
\citep{2001ApJ...551..807D}. The
thermal energy in the gas per unit volume of the disk is:
\begin{equation}\label{eq-e-th-gas}
e_{\mathrm{th,g}} = c_{V,g}\,\rho_g\,T_g \comma
\end{equation}
\revised{with the specific thermal heat capacity of the gas given by}
\begin{equation}\label{eq-def-cv-gas}
c_{V,g} = \frac{k_B}{(\gamma-1)\mu m_u} \comma
\end{equation}
where $k_B$ is the Boltzmann constant, $\mu\simeq 2.3$ is the mean molecular
weight of the gas in units of the atomic unit mass $m_u$, and $\gamma$ is the
ratio of specific heats. The total thermal
energy density is the sum of the two
\begin{equation}
e_{\mathrm{th}} = e_{\mathrm{th,g}} + e_{\mathrm{th,\smalldust}} \fullstop
\end{equation}
For a small-grain dust-to-gas ratio smaller than or equal to 0.01 we can safely approximate
this as $e_{\mathrm{th}} = e_{\mathrm{th,g}}$. The optically thin radiative cooling time is
then
\begin{equation}\label{eq-cooling-time-rad}
  t_{\mathrm{cool,thin}}^{\mathrm{rad}} = \frac{e_{\mathrm{th,g}}}{q_{\mathrm{cool,\smalldust}}} \comma
\end{equation}
assuming $T_{\smalldust}=T_g=\tmid$. \revised{In the optically thin limit this then becomes}
\begin{equation}\label{eq-cooling-time-rad-thin}
  t_{\mathrm{cool,thin}}^{\mathrm{rad}} =
  \frac{c_{V,g}}{4\sigma_{\mathrm{SB}}\kappa_P(T)}
  \frac{\rho_g}{\rho_{\smalldust}}\frac{1}{T^3} \fullstop
\end{equation}

\revised{However, what we need for the analysis of the
VSI is the {relaxation time}, which is shown in Appendix
\ref{app-thermal-relaxation-time-thin}, to be}
\begin{equation}\label{eq-relax-time-rad}
  t_{\mathrm{relax,thin}}^{\mathrm{rad}} = \frac{1}{4+b}\,t_{\mathrm{cool,thin}}^{\mathrm{rad}} \comma
\end{equation}
\revised{where}
\begin{equation}
b=\frac{d\ln\kappa_P(T)}{d\ln T} \simeq 1.7 \comma
\end{equation}
\revised{where the $1.7$ is valid for the opacity model of
Eq.~(\ref{eq-fitting-formula-planck}) of Appendix
\ref{app-planck-mean-opacity}.}

\revised{So far we have not included optical depth effects, and have therefore considered
the most VSI-friendly scenario. Optical depth effects can only increase the
relaxation time, not shorten it. We are primarily interested in the regions that
are spatially resolvable with ALMA, meaning we are interested in $r\gtrsim 10$
au. The optical depth of the disk to its own radiation is moderate to low in
these outer regions. Optical depth effects are therefore not expected to play a
large role in these regions. But it is not a major effort to include them.
In Appendix \ref{app-thermal-relaxation-time-thick} we discuss the relaxation
time scale in the optically thick regime, and write it as
$t_{\mathrm{cool,thick}}^{\mathrm{rad}}$ given by
Eq.~(\ref{eq-trelax-thick}).}

\revised{Finally, we have to account for} the time it takes to transfer
heat between the gas and the dust, $t_{\dustgas}$. \revised{This will play
  a big role for disks around bright stars such as Herbig Ae/Be stars,
  where it will be the limiting factor of the radiative the cooling.}
In Appendix \ref{app-thermal-coupling-dust-gas} we give an expression for
$t_{\dustgas}$.

\revised{We estimate the combined cooling time scale to be the
  sum of all three time scales\footnote{A python tool to compute the relaxation
  time scale is available at \url{https://github.com/dullemond/ppdiskcoolcalc}}:}
\begin{equation}\label{eq-cooling-time-netto}
t_{\mathrm{relax}} = t_{\mathrm{relax,thin}}^{\mathrm{rad}} + t_{\mathrm{relax,thick}}^{\mathrm{rad}} + t_{\dustgas} \comma
\end{equation}
\revised{which gives a smooth transition between regions, and ensures that the
  limiting factor determines the actual relaxation time.}

In Fig.~\ref{fig-tcool-vsi} this \revised{relaxation} time is shown for the fiducial disk
model of Appendix \ref{app-fiducial-disk-model}, for small-grain dust-to-gas
ratios of $Z_{\mathrm{small}}=10^{-2}$ (no depletion, i.e., $\coag=0$), for
$Z_{\mathrm{small}}=10^{-3}$ (a factor of 10 depletion of small dust grains,
i.e., $\coag=0.9$), and for $Z_{\mathrm{small}}=10^{-4}$ (a factor of 100
depletion of small dust grains, i.e., $\coag=0.99$).  The dotted lines represent
$t_{\mathrm{relax,thin}}^{\mathrm{rad}} + t_{\dustgas}$, i.e., without
optical depth effects.

\begin{figure}
  \centerline{\includegraphics[width=0.5\textwidth]{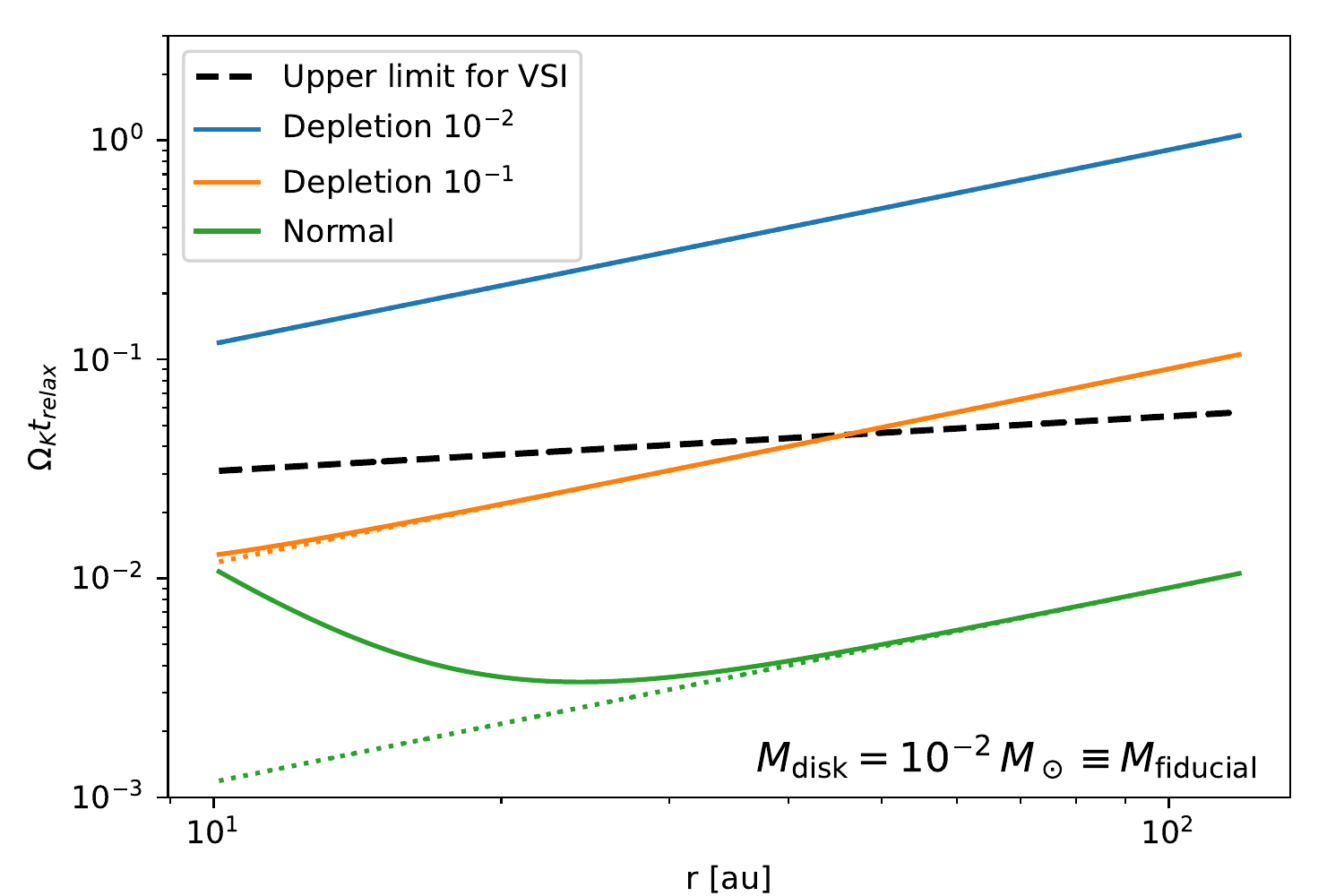}}
  \caption{\label{fig-tcool-vsi}\revised{Relaxation} time (Eq.~\ref{eq-cooling-time-netto}) of the disk in units of the
    Kepler time $\Omega_K^{-1}$ for \revised{three} small-grain dust-to-gas ratios:
    \revised{Normal ($Z_{\mathrm{small}}=10^{-2}$), depleted by a factor of $10^{-1}$ ($Z_{\mathrm{small}}=10^{-3}$)
    and depleted by a factor of $10^{-2}$ ($Z_{\mathrm{small}}=10^{-4}$)}. Dotted
    lines: optically thin approximation (Eq.~\ref{eq-relax-time-rad}). Solid
    lines: including optical depth effects.  In dashed black: upper limit to
    the \revised{relaxation} time (Eq.~\ref{eq-vsi-upper-limit-tcool}) for which VSI is
    operational is shown. The disk and stellar parameters are those of the star
    Oph 163131 (the fiducial model of Appendix \ref{app-fiducial-disk-model}).}
\end{figure}

In Fig.~\ref{fig-tcool-vsi-10xlowermass}
the same is shown for a 10 times lower disk mass, both in dust and in gas.
Because of the lower optical depth, the solid curves are now closer to
the optically thin estimates.

One can see that for both the fiducial model and for the 10$\times$ lower mass
disk, $\Omega_K t_{\mathrm{relax}}$ is well below unity, justifying the locally isothermal
appoximation for most applications. However, for the VSI to be operational, the
\revised{relaxation} time has to be below the limit given in
Eq.~(\ref{eq-vsi-upper-limit-tcool}), shown with the thick dashed line in
Figs.~\ref{fig-tcool-vsi} and \ref{fig-tcool-vsi-10xlowermass}.

\revised{For the fiducial model, with small-grain dust-to-gas ratio of $10^{-2}$, the
thermal relaxation time scale is everywhere below this limit, meaning
that the disk is prone to the VSI everywhere. However, if dust
coagulation converts, say, 90\% of the small grains into large dust aggregates
(a depletion of $10^{-1}$, or $\coag=0.9$), leading to a small-grain dust-to-gas ratio of
$Z_{\mathrm{small}}=10^{-2}(1-\coag)=10^{-3}$, then the $\Omega_K t_{\mathrm{relax}}$ is
above the threshold value for $r\gtrsim 50\;\mathrm{au}$. If coagulation converts
99\% of the small grains (a depletion of $10^{-2}$, or $\coag=0.99$), then the
curve is everywhere well above the threshold, and the entire disk is VSI-stable.}

\begin{figure}
  \centerline{\includegraphics[width=0.5\textwidth]{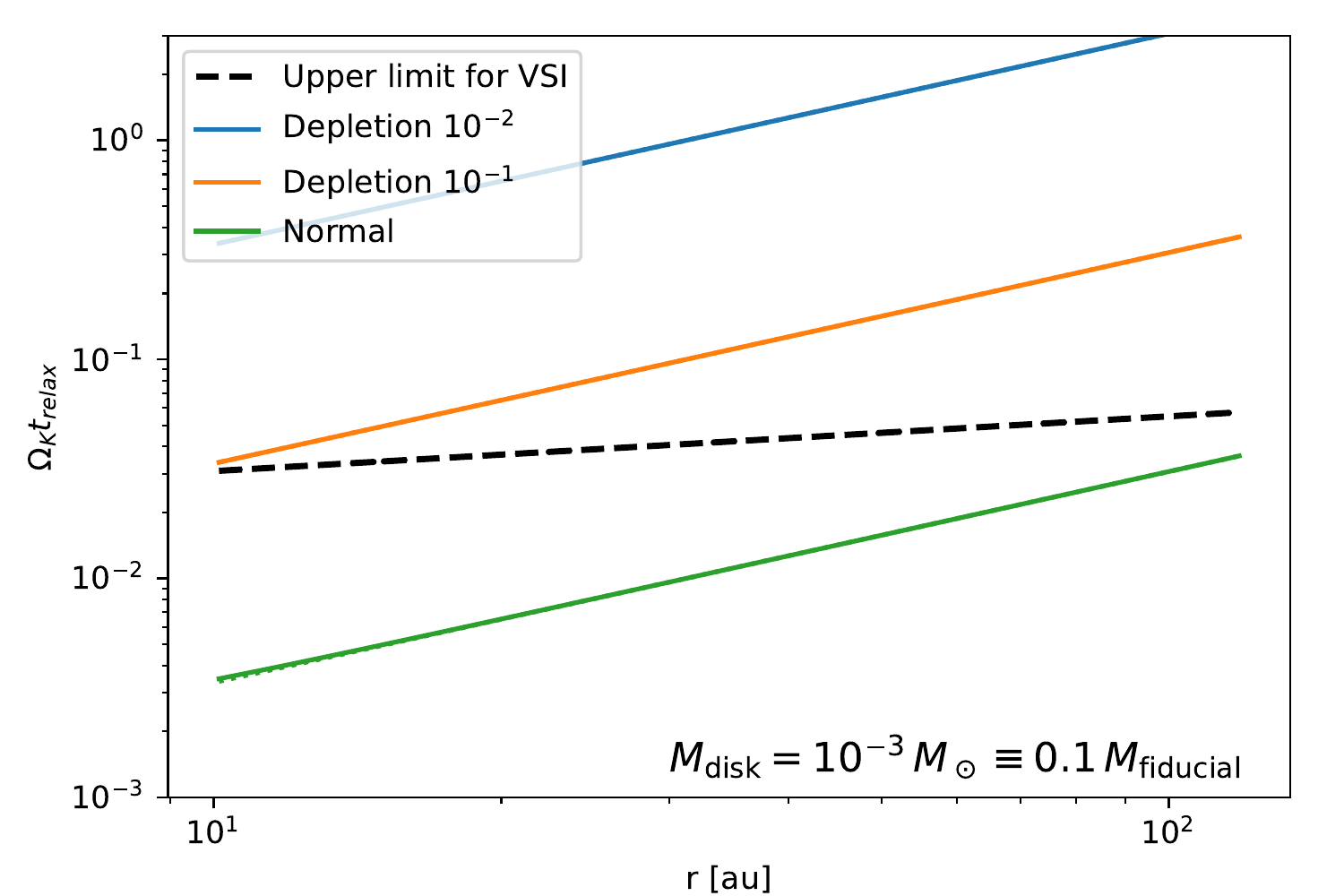}}
  \caption{\label{fig-tcool-vsi-10xlowermass}As Fig.~\ref{fig-tcool-vsi}, but now for
  10x lower mass as the fiducial model, both in dust and gas.}
\end{figure}

If we redo our analysis for a brighter star, say a Herbig Ae star, then the disk
will be warmer due to the stronger irradiation. This will lower the cooling
times and thus make the disk more susceptible to the VSI. We show the resulting
cooling time scales for a Herbig Ae star of $M=2.4\,M_{\odot}$ and
$L=50\,L_{\odot}$, with otherwise the same disk parameters, in
Fig.~\ref{fig-tcool-vsi-herbigae}. Indeed, the cooling time for a normal
dust-to-gas ratio is substantially shorter. A depletion of small grains of
a factor of 10 is not sufficient, but a factor of 100 will, again, make
the disk stable against the VSI in most of the disk.

\begin{figure}
  \centerline{\includegraphics[width=0.5\textwidth]{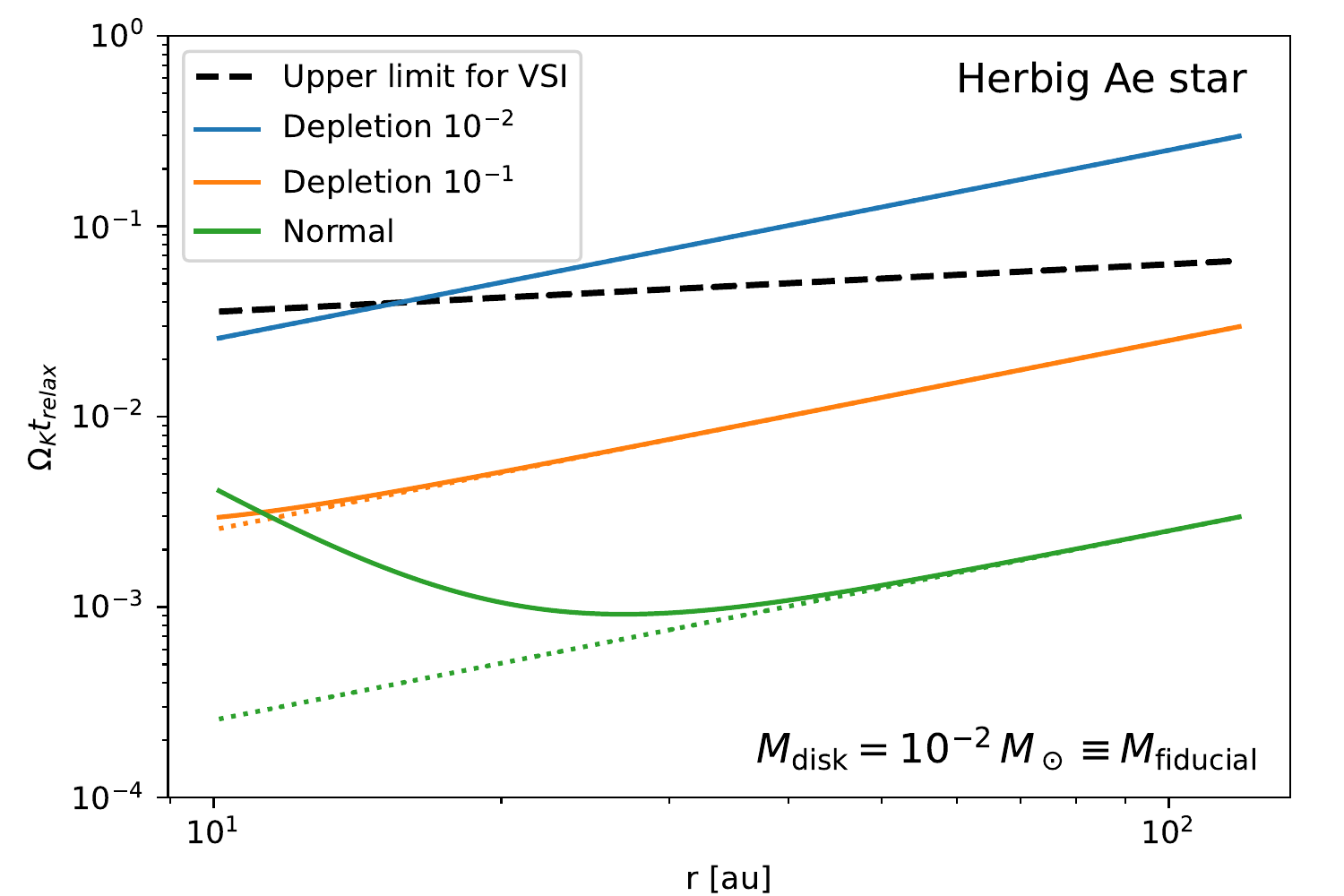}}
  \caption{\label{fig-tcool-vsi-herbigae}As Fig.~\ref{fig-tcool-vsi}, but now
    for a disk around a Herbig Ae star with $M=2.4\,M_{\odot}$ and
    $L=50\,L_{\odot}$.}
\end{figure}

A depletion of small dust grain by a factor of 10 or even 100 due to coagulation
is not extreme. The fact that most protoplanetary disks look ``fat''
(geometrically vertically extended) in optical and near-infrared observations is
not evidence of a lack of coagulation. \revised{In Appendix
  \ref{sec-opt-appearance-depletion} we quantify this by computing the optical
  appearance of our fiducial disk at a wavelength of $\lambda=0.8\,\mu$m for
  various degrees of small-grain depletion. These images show that the typical
  appearance of the disk, with its two bright layers separated by a dark lane,
  is retained even at large degrees of depletion.}  The required amount of dust
coagulation to inhibit the VSI is therefore within the observational
constraints.

\section{Discussion}\label{sec-discussion}

\subsection{Earlier work on the stirring of large grains by the VSI}
The extreme effectiveness of the VSI to stir up even large dust aggregates to
high elevations above the midplane is not a new result, and has been noted by
several previous authors. \citet{2017ApJ...850..131F, 2020ApJ...897..155F}
presented detailed 3D radiation hydrodynamical models of protoplanetary disks
with dust particle dynamics. They show that 0.1 mm and 1 mm dust grains achieve
greater elevations above the midplane than expected from isotropic turbulence.
Similar conclusions were also made by \citet{2022A&A...658A.156L}, who show the
dependency of this effect on Stokes number $\mathrm{St}_0$ and
vertically integrated dust-to-gas ratio $Z_{\bigdust}$, although they focus on smaller
values of the Stokes number than we explore in this paper. However, we put this
into context with recent observational evidence of the extreme vertical
geometrical thinness of the big-grain dust layers in (most?) protoplanetary
disks.

\subsection{Effect of dust traps}
The fact that our models are for disks without dust traps limits the
applicability of the results. If the large dust grains in the outer regions of
protoplanetary disks remain at those large distances because they are trapped,
then it is rather natural to get $Z_{\bigdust}\gg 0.01$ in these traps because
all the dust elsewhere will radially drift into these traps, enhancing $Z$
there. \revised{As argued by \citet{2019MNRAS.485.5221L} and
  \citet{2022A&A...658A.156L}, this then could naturally push
  $Z_{\bigdust}\gtrsim 0.02\cdots0.05$ which, according to their simulations,}
strongly suppresses the VSI.

\revised{The fact that the upper limit of $Z_{\bigdust,VSI}$ for the VSI lies
  around the same value as the lower limit of $Z_{\bigdust,SI}$ for streaming
  instability \citep{2017ApJ...839...16C} leads to an interesting
  speculation. It was shown by \citet{2019ApJ...884L...5S}, using an argument
  related to that of Section \ref{section-case-against-St-big}, that
  $Z_{\bigdust,SI}$ coincides with an optical depth at millimeter wavelengths of
  order unity, as appears to be observed in ALMA observations. They argue that
  dust traps attract more and more dust, until $Z_{\bigdust}$ reaches
  $Z_{\bigdust,SI}$, at which point $Z_{\bigdust}$ stabilizes: Any further dust
  added to the trap will be converted into planetesimals by the streaming
  instability, keeping $Z_{\bigdust}=Z_{\bigdust,SI}$. If
  $Z_{\bigdust,SI}>Z_{\bigdust,VSI}$, then this self-regulating system naturally
  keeps the disk VSI-stable. This could be another natural explanation for
  the lack of VSI, but this requires more detailed study of the
  combined VSI+SI, as in \citet{2020A&A...635A.190S}. So at this point,
  this is merely a speculative idea.}

\subsection{Effect of small but non-zero background turbulence}
\revised{It was noted by \citet{2013MNRAS.435.2610N} that the VSI is also damped
  if the viscosity parameter of the disk $\alpha\gtrsim 4\times 10^{-4}$, i.e.,
  typically when the disk is turbulent due to the magnetorotational instability
  (MRI).  While MRI turbulence will also stir up large grains away from the
  midplane, it is far less effective than the VSI. And so, somewhat
  paradoxically, the existence of weak, but no-zero, turbulence might, by
  inhibiting the VSI, allow large grains to settle to a thinner layer than is
  the case for a non-turbulent disk.}

\subsection{3D effects}
\revised{Since our models are 2D in $(r,z)$, we cannot treat any potential
  non-axisymmetric modes, such as the formation of long-lived vortices
  \citep{2022A&A...658A.156L}. The main effect of the VSI acts in the $(r,z)$
  directions, however, and is not dependent on the $\phi$-direction. Any 3D
  effects on the large-scale may affect the observational appearance of the
  disk, but will likely not affect our conclusion that the flat disks seen with
  ALMA are incompatible with the VSI.}

\subsection{Small-scale modes}
\revised{The VSI may operate on smaller scales than we can model with our global
  models, for example, via the parametric instability mechanism described by
  \citet{10.1093/mnras/stac279}. This can have consequences for the dust
  dynamics and dust growth. If the VSI operates on the large scales as explored
  in this paper, the observational consequences will be dominated by these
  large-scale motions, even if smaller-scale motions are superposed on them. However,
  as shown in \citep{10.1093/mnras/stac279}, the small-scale motions excited by
  the larger ones act as an energy sink to the large-scale motions. The
  long-term saturation state of the large-scale VSI modes may therefore depend
  on the very small-scale motions that require super-high spatial resolution to
  resolve.  \citep{10.1093/mnras/stac279} cite a resolution of 300 grid cells
  per scale height, which is out of reach for global simulations. These
  considerations show that it remains to be explored to which degree the
  VSI or any other instabilities are inhibited if a protoplanetary disk is
  observed to have a very flat midplane dust layer, and what this means for the
  implied conditions in the disk.}

\subsection{Uncertainties in the thermal relaxation time}
\revised{Our estimates of the thermal relaxation time suffer from some
  uncertainties. First, they are very sensitive to the disk temperature profile
  $T(r)$. This is because radiative cooling goes as $\propto T^4$. But in
  practice this sensitivity is not so severe, because any uncertainty in the
  irradiation $q_{\mathrm{heat}}\propto L_{*}$ of the disk only enters the
  temperature as $T\propto q_{\mathrm{heat}}^{1/4}\propto L_{*}^{1/4}$. It
  does show, however, that for disks around Herbig Ae stars the relaxation times are
  smaller than for T Tauri stars.}

\revised{A much more critical uncertainty is the dust opacity at long
  wavelengths. The popular \citet{1994ApJ...427..987B} opacity represents a
  relatively low estimate, lower than what we use in this paper. This opacity
  model was used by \citet{2015ApJ...811...17L} and \citet{2019ApJ...871..150P}
  for their relaxation time estimates, which leads to less favorable conditions
  for the VSI than our estimates. \citet{2017A&A...605A..30M} use the opacity
  model of \citet{2003A&A...410..611S}, which is, for $\tmid<100$ K, very
  similar to that of \citet{1994ApJ...427..987B}. The factor of $\sim$1.5
  difference can mostly by explained by the use of a different dust-to-gas
  ratio, so that the dust opacities are more or less the same.  In contrast,
  \citet{2021ApJ...914..132F} use a simple analytic opacity model
  \citep[see][]{1997MNRAS.291..121I}, which is higher than what we use in this
  paper, and leads to more favorable conditions for the VSI than our estimates
  (although this effect is limited by the dust-gas coupling time scale that
  becomes the limiting factor). The comparison of these opacities is shown in
  Fig.~\ref{fig-kappa-mean}. The opacity at the far-infrared and submillimeter
  of the dust in protoplanetary disks is notoriously uncertain, and the real
  opacity is likely somewhere between these two extremes.}

\revised{Another major uncertainty is the dust-to-gas ratio. In our analysis we
  kept this constant at a value of $Z=0.01$, although we allowed coagulation to
  convert the small grain population (responsible for the radiative cooling)
  into a big grain population (responsible for the dust observed with ALMA).
  However, these big grains can radially drift, leading to a reduction of the
  dust-to-gas ratio in the outer regions. However, since the big grains do not
  contribute to the cooling, this does not affect our analysis. What matters is
  only the small grain abundance $Z_{\smalldust}$. Dust coagulation can reduce
  $Z_{\smalldust}$. What subsequently happens to $Z_{\bigdust}$ is irrelevant
  for the estimation of $t_{\mathrm{relax}}$.}

\subsection{Non-flat rings}
It should be noted that there are observed protoplanetary disks for which one or
more of the rings do  not appear to be geometrically very thin. For
instance, \citet{2021ApJ...912..164D} conclude, after detailed radiative
transfer modeling, that the inner dust ring of HD 163296 is, in fact, likely to
be vertically extended to about a gas pressure scale height. One interpretation
of this could be that in this ring the cooling rate is, in fact, not
\revised{sufficiently} reduced by grain growth, so that the VSI is operational.
\revised{From Fig.~\ref{fig-tcool-vsi-herbigae} it can be seen that for a
  small-grain depletion somewhere around $3\times 10^{-2}$, the inner disk regions
  remain unstable to the VSI while the outer disk regions stabilize against the
  VSI, which might explain why the inner ring of HD 163296 is vertically more
  extended than the outer one. A similar point was made by
  \citet{2021ApJ...914..132F}. The disk of HD 163296 is fairly dim at those
  wavelengths \citep{2022A&A...658A.137G}, which does seem to point to
  substantial small-grain depletion.}


\section{Conclusions}\label{sec-conclusion}
In this paper we show that the geometrically very thin midplane layers of dust
observed in many protoplanetary disks, most strikingly shown in a recent paper
by \citet{2022arXiv220400640V}, are evidence that the 
VSI is not operational in these outer disk regions. Dust particles with
dimensionless stopping times $\St\lesssim 1$ would be stirred up by the VSI to a
substantial fraction of the gas pressure scale height, even for large particles
with $\St\simeq 0.1-1$. Only for even larger particles, with $\St\gtrsim 1$,
does the dust layer remain largely unaffected by the VSI. But we show that to
have such a layer being \revised{marginally} optically thick
\revised{($\tau\gtrsim 0.3$)} at ALMA wavelengths, as seems to be the case for
many such rings, this requires a vertically integrated dust-to-gas ratio
$Z_{\bigdust}\gtrsim 0.08\cdots 0.16$, which \revised{we consider} an unlikely
scenario.

Damping or inhibiting the VSI in the outer regions of a protoplanetary disk can
be due to an enhanced vertically integrated dust-to-gas ratio of
$Z_{\bigdust}\gtrsim 0.02\cdots 0.04$, \revised{as shown by
  \citet{2019MNRAS.485.5221L} and \citet{2022A&A...658A.156L}, or due to a
  modest background turbulence \citep{2013MNRAS.435.2610N}.}

We show that another possible explanation is that dust
coagulation has converted more than 90\% of the small grains in the disk into
big grains (likely the ones that make up the midplane dust layer). In that case,
the gas cannot cool fast enough through the thermal emission of the small
grains, and the VSI is inhibited, as shown by
\citet{2015ApJ...811...17L}. Small-grain dust depletion of more than
\revised{90\%} by coagulation \revised{($\coag=0.9$)} is reasonable during the
life time of these disks \citep{2005ApJ...625..414T, 2005A&A...434..971D,
  2010A&A...513A..79B}, and remains consistent with the SEDs of these disks
\citep{2004A&A...417..159D}.


Our conjecture is thus that protoplanetary disks that show geometrically thin
disks in ALMA observations are stable against the VSI.

\begin{acknowledgements}
  \revised{We thank Mario Flock for useful discussions that helped us find an error
  in an earlier version of this manuscript. We also thank the anonymous
  referee for insightful comments and suggestions that allowed us to improve the paper.}
  This research was supported by the Munich Institute for Astro- and Particle
  Physics (MIAPP) which is funded by the Deutsche Forschungsgemeinschaft (DFG,
  German Research Foundation) under Germany´s Excellence Strategy – EXC-2094 –
  390783311. We acknowledge funding from the DFG research group FOR 2634
  ``Planet Formation Witnesses and Probes: Transition Disks'' under grant DU
  414/22-2 and KL 650/29-2, 650/30-2. AZ is supported by STFC grant
  ST/P000592/1. The authors acknowledge support by the High Performance and
  Cloud Computing Group at the Zentrum f\"ur Datenverarbeitung of the University
  of T\"ubingen, the state of Baden-W\"urttemberg through bwHPC and the German
  Research Foundation (DFG) through grant no INST 37/935-1 FUGG.
\end{acknowledgements}

\bibliographystyle{apj}
\bibliography{ms}

\appendix

\section{Fiducial disk model}
\label{app-fiducial-disk-model}
To be able to make estimates of the numbers for this paper, we choose as
standard star the star Oph 163131, for which we take the parameters from the
paper \citet{2022arXiv220400640V}: $M_{*}=1.2\,M_\odot$, $R_{*}=1.7\,R_\odot$,
$T_{*}=4500\,\mathrm{K}$, which leads to $L_{*}\simeq L_\odot$. As an estimate
of the disk midplane temperature we use a simple flaring disk estimate:
\begin{equation}
\tmid(r) = \left(\frac{\tfrac{1}{2}\varphi L_{*}}{4\pi r^2\sigma_{\mathrm{SB}}}\right)^{1/4} \comma
\end{equation}
where $\varphi$ is the flaring irradiation angle (radiative incidence angle),
which we take, somewhat arbitrarily, as $\varphi=0.05$. Usually this value gives
temperatures that are reasonably consistent with observational measurements.
From this midplane temperature the gas pressure scale height $h_P(r)$ can be
computed through $h_p=c_s/\Omega_K$, with $c_s=\sqrt{k_B\tmid/\mu m_u}$ the
isothermal sound speed and $\Omega_K=\sqrt{GM_{*}/r^3}$ the Kepler frequency.
The disk is flaring with $d\ln(h_p/r)d\ln(r)=0.25$. 

We choose as reference radius $r_0=100\,\mathrm{au}$. With the above parameters,
the temperature at this reference radius is 16 K, and
the gas pressure scale height is $h_p(r_0)/r_0=0.0732$.
The radial distribution of gas and dust mass are assumed to follow a powerlaw
\begin{eqnarray}
\Sigma_g(r) &=& \Sigma_{g,100} \left(\frac{r}{100\,\mathrm{au}}\right)^{-1} \comma\\
\Sigma_{\mathrm{\smalldust}}(r) &=& \Sigma_{\smalldust,100} \left(\frac{r}{100\,\mathrm{au}}\right)^{-1} \comma\\
\Sigma_{\bigdust}(r) &=& \Sigma_{\bigdust,100} \left(\frac{r}{100\,\mathrm{au}}\right)^{-1} \comma
\end{eqnarray}
where $g$ stands for gas, $\smalldust$ for small grains and $\bigdust$ for big grains.
The normalization factors $\Sigma_{x,100}$ can be expressed in terms
of the disk mass
\begin{equation}
\begin{split}
  M_{x} &= 2\pi \int_{r_{\mathrm{in}}}^{r_{\mathrm{out}}} \Sigma_{x}(r) r^2dr \comma\\
  &= 7\times 10^{-5}\,\Sigma_{x,100} \left(\frac{r_{\mathrm{out}}}{\mathrm{au}}-\frac{r_{\mathrm{in}}}{\mathrm{au}}\right)\;M_{\odot} \fullstop
\end{split}
\end{equation}
For the direct comparison to Oph 163131, we take
$r_{\mathrm{in}}=10\,\mathrm{au}$, $r_{\mathrm{out}}=120\,\mathrm{au}$, for
which we obtain
\begin{equation}
M_x = 7.78\times 10^{-3}\,\Sigma_{x,100}\;M_{\odot} \fullstop
\end{equation}
For the fiducial disk model we choose $M_{g}=0.01\,M_{\odot}$,
i.e., $\Sigma_{g,100}=1.29\,\mathrm{g}/\mathrm{cm}^2$.

For the simulations of the VSI we take $r_{\mathrm{out}}=500\,\mathrm{au}$
(keeping $\Sigma_{g,100}$ fixed), to ensure that around $r=r_0$ there are no
boundary effects.

The initial vertical structure of the gas at the start of the hydrodynamic
simulation is set up as follows. We use the equilibrium state derived in
\citet{2013MNRAS.435.2610N} with $p=-2.25$ and $q=-0.5$ such that $\Sigma\propto
1/R$.  The velocity field is seeded with noise with an amplitude of 1\% local
$c_s$ to kickstart the VSI.

\section{Dust opacity model}
\label{app-dust-opacity-model}
Both the small dust grains that are suspended in the gas and the large dust
grains that make up the midplane layer seen by ALMA are assumed to consist of
the same \revised{refractory} material: 0.87 mass fraction of pyroxene with 70\%
magnesium, and 0.13 mass fraction of amorphous carbon. This is the DIANA
standard mixture \revised{\citep{2016A&A...586A.103W}}.  In addition we add a
mantel of water ice such that the water ice contributes 20\% of the total mass
\citep{2016Natur.530...63P}. This reduces the \revised{mass} fractions of pyroxene and carbon
to 0.696 and 0.104, respectively. We assume a porosity of $p=$0.25. The material
density of a dust grain made up of this material is
$\rho_s=1.48\,\mathrm{g/cm}^3$.

To compute the opacities we use the publicly available code {\small\tt
  optool}\footnote{\url{https://github.com/cdominik/optool}}
\citep{2021ascl.soft04010D}. We use the ``Distribution of Hollow Spheres (DHS)''
method \citep{2005A&A...432..909M} with $f_{\mathrm{max}}=0.8$. The
opacities for different grain sizes are shown in Fig.~\ref{fig-kappa-lambda}.
Whereever we plot or use the scattering opacity, we use the effective
scattering opacity given by
\begin{equation}\label{eq-effective-scattering-kappa}
\kappa_\nu^{\mathrm{scat,eff}} = (1-g_\nu)\,\kappa_\nu^{\mathrm{scat}} \comma
\end{equation}
where $g_\nu$ is the scattering anisotropy coefficient. The factor $(1-g_\nu)$
is a simple way to account for anisotropy \citep{2009A&A...497..155M}. This
effective scattering opacity gives a better estimate of the importance of
scattering compared to absorption, because it weighs strongly forward scattering
($g_\nu\rightarrow 1$) less than isotropic scattering ($g_\nu\rightarrow 0$).

\begin{figure}
  \centerline{\includegraphics[width=0.5\textwidth]{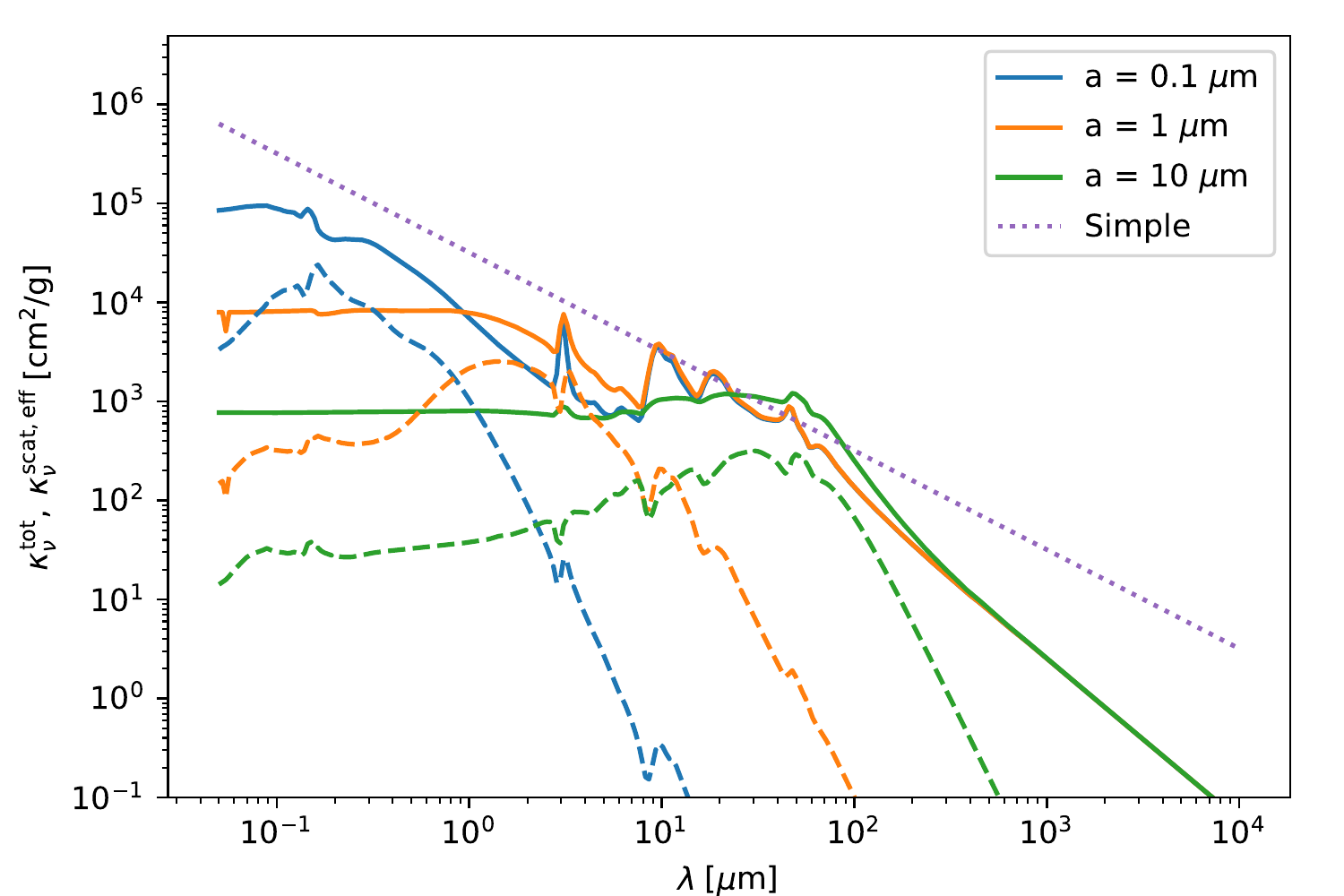}}
  \caption{\label{fig-kappa-lambda}Dust grain opacity model used for this
    study. Shown here is the opacity as a function of wavelength for three grain
    sizes \revised{in units of cm$^2$ per gram of dust}.
    Solid line: total effective opacity $\kappa_\nu^{\mathrm{tot}}
    =\kappa_\nu^{\mathrm{abs}}+\kappa_\nu^{\mathrm{scat,eff}}$; dashed line:
    effective scattering opacity
    $\kappa_\nu^{\mathrm{scat,eff}}=(1-g_\nu)\kappa_\nu^{\mathrm{scat}}$.
    \revised{The dotted line represents the simple
      $\kappa_\lambda=3\pi/2\rho_s\lambda$ opacity model of
      \citet{1997MNRAS.291..121I} in the vanishingly small grain
      size limit.}}
\end{figure}

\revised{Also plotted in Fig.~\ref{fig-kappa-lambda}, for comparison, is a
  simple analytic dust opacity model proposed by \citet{1997MNRAS.291..121I}, in
  the limit of vanishingly small grain size. This model, and the version for
  finite grain sizes ($\kappa_\lambda=(\pi a^2/m)\min(1,2\pi a/\lambda)$, where
  $m$ is the grain mass and $a$ the grain radius), is often used in the
  literature because of its simplicity. It is also used in
  \citet{2021ApJ...914..132F} to estimate the thermal relaxation time scale in
  protoplanetary disks.}

\section{Rosseland and Planck mean opacity of the small dust grains}
\label{app-planck-mean-opacity}
For the opacity model of Appendix \ref{app-dust-opacity-model} we compute here
the Rosseland and Planck mean opacities. The Planck mean opacity was already
defined in the main text (Eq.~\ref{eq-define-planck-mean}). The Rosseland mean
opacity is defined as
\begin{equation}\label{eq-define-rosseland-mean}
  \kappa_R(T_{\smalldust}) = \frac{\int_0^\infty (\partial B_\nu(T_{\smalldust})/\partial T) d\nu}{\int_0^\infty (\partial B_\nu(T_{\smalldust})/\partial T)/\kappa^{\mathrm{tot}}_{\nu,\smalldust}d\nu} \comma
\end{equation}
where
\begin{equation}\label{eq-total-kappa}
\kappa^{\mathrm{tot}}_{\nu,\smalldust} = \kappa^{\mathrm{abs}}_{\nu,\smalldust} + \kappa^{\mathrm{scat,eff}}_{\nu,\smalldust} \comma
\end{equation}
where the effective scattering opacity $\kappa^{\mathrm{scat,eff}}_{\nu,\smalldust}$ is
defined as in Eq.~(\ref{eq-effective-scattering-kappa}).

\begin{figure*}
  \centerline{\includegraphics[width=0.5\textwidth]{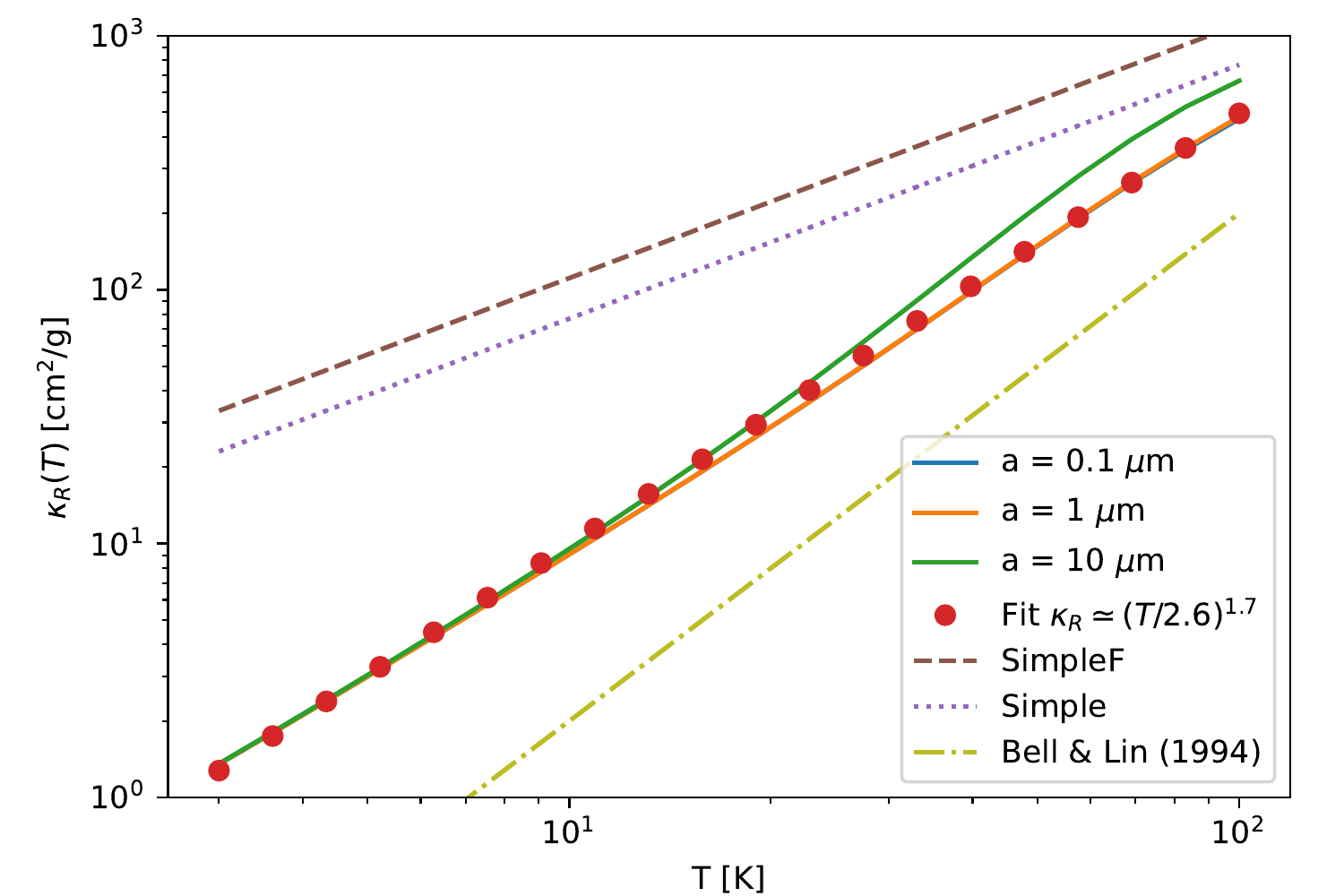}
  \includegraphics[width=0.5\textwidth]{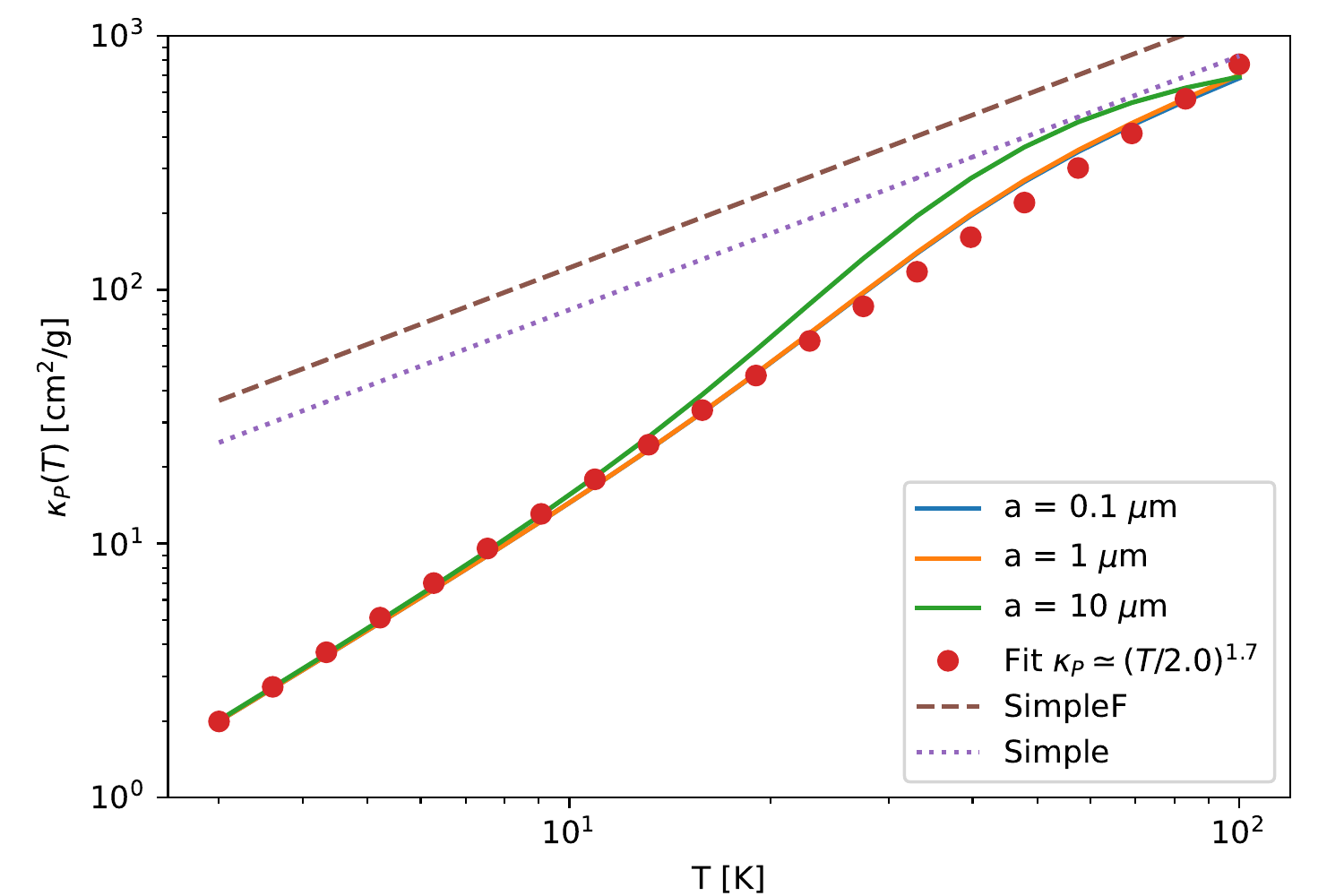}}
  \caption{\label{fig-kappa-mean}Small dust grain mean opacities as a
    function of temperature \revised{for three grain sizes}.  Left: Rosseland
    mean. Right: Planck mean. In the left panel the often-used opacity of
    \citet{1994ApJ...427..987B} is plotted for comparison, \revised{which is
      used by \citet{2015ApJ...811...17L}, \citet{2019ApJ...871..150P} and many
      other works}. \revised{\citet{2017A&A...605A..30M} use the dust opacity model
      of \citet{2003A&A...410..611S}, which is, however, very similar to that
    of \citet{1994ApJ...427..987B} in this temperature range.}
    \revised{Also for comparison, the top two curves in each
      panel show the mean opacity for the simple
      $\kappa_\lambda=3\pi/2\rho_s\lambda$ small-grain opacity model used by
      \citet{2021ApJ...914..132F}, for our value of the material density
      $\rho_s=1.48\;\mathrm{g}\,\mathrm{cm}^{-3}$ (Simple) and their value
      of $\rho_s=1.00\;\mathrm{g}\,\mathrm{cm}^{-3}$ (SimpleF).}}
\end{figure*}

The resulting mean opacities are shown in Fig.~\ref{fig-kappa-mean}. As one can
see, for small enough grains and small enough temperatures, the Planck mean
opacity of the dust can be well approximated for $T\lesssim 100\,\mathrm{K}$ and
$a\lesssim 10\,\mu\mathrm{m}$ by the following fitting formula:
\begin{equation}\label{eq-fitting-formula-planck}
\kappa_P^{\mathrm{fit}}(T) = \left(\frac{T}{2.0\,\mathrm{K}}\right)^{1.7}\; \frac{\mathrm{cm}^2}{\mathrm{g}} \comma
\end{equation}
where $\mathrm{g}$ is to be interpreted as gram of small-grain dust. The symbol
$K$ is the unit of Kelvin. Likewise
the Rosseland mean opacity can be approximated as
\begin{equation}\label{eq-fitting-formula-rosseland}
\kappa_R^{\mathrm{fit}}(T) = \left(\frac{T}{2.6\,\mathrm{K}}\right)^{1.7}\; \frac{\mathrm{cm}^2}{\mathrm{g}} \comma
\end{equation}
For comparison, the dusty part of the \citet{1994ApJ...427..987B} opacity is
$\kappa_{R,\mathrm{BellLin}}(T)=2\times
10^{-2}\,(T/\mathrm{K})^2\;\mathrm{cm}^2/\mathrm{g}\simeq
(T/7.07\,\mathrm{K})^2\;\mathrm{cm}^2/\mathrm{g}$, where we assume a dust-to-gas
ratio of 0.01. \revised{The equivalent Planck-mean opacity for Bell \& Lin would
  be
  $\kappa_{P,\mathrm{BellLin}}(T)=(T/4.58\,\mathrm{K})^2\;\mathrm{cm}^2/\mathrm{g}$.
  This means that our opacity model is more favorable to the onset of the VSI
  than the Bell \& Lin opacity, implying that in our analysis we need to reduce
  the small-grain dust more strongly than the analysis done by
  \citet{2015ApJ...811...17L} and \cite{2019ApJ...871..150P} to suppress the
  VSI.}

\revised{The primary reason why our opacity model exceeds that of Bell \& Lin is
  the amorphous carbon mixed into the composition, which is responsible for the
  ``antenna effect'', which strongly enhances the long-wavelength opacity. If
  the carbon would be, instead, in the form of organics, this effect
  would not be seen. The 25\% porosity also increases the opacity a bit. If we
  use pure pyroxene without porosity, our opacity would exceed that of Bell \&
  Lin by only 18\%, which is well within the uncertainty of the dust-to-gas
  ratio we used to convert the Bell \& Lin total opacity to a dust-only
  opacity. We refer to \citet{2016A&A...586A.103W} for details on the role of
  carbon.}

\section{Thermal relaxation time scale: The optically thin case}
\label{app-thermal-relaxation-time-thin}
\revised{The thermal relaxation time scale is not identical to the cooling time
scale. The radiative cooling time scale $t_\mathrm{cool}^{\mathrm{rad}}$ is
given by Eq.~(\ref{eq-cooling-time-rad}), assuming $T_{\smalldust}=T_g=\tmid$,
$e_{\mathrm{th}}=e_{\mathrm{th,g}}$, $\rho_d\ll \rho_g$, and
$q_{\mathrm{cool}}=q_{\mathrm{cool,small}}$. In the optically thin limit this
becomes Eq.~(\ref{eq-cooling-time-rad-thin}). This is the time scale of cooling
if the heating were to be suddenly switched off.}

\revised{\revisedd{However, the condition of \citet{2015ApJ...811...17L} was derived for
  Newtonian cooling (linear in temperature), not for radiative cooling
  (proportional to $T^4$). This means that for the VSI we require the 
    thermal relaxation time, which is the time scale on which a small deviation
  $\delta T$ from thermal equilibrium would exponentially decay}
  \citep{2017A&A...605A..30M,2019ApJ...871..150P}. Local thermodynamic
  equilibrium implies $q_{\mathrm{cool}}=q_{\mathrm{heat}}$, where
  $q_{\mathrm{heat}}$ is, in our case, the irradiation of the disk by the
  star. The time-dependent equation for the thermal energy in an optically thin
  disk is then
\begin{equation}\label{eq-timedep-cooling-eq}
\frac{de_{\mathrm{th,g}}(T)}{dt} = q_{\mathrm{heat}} - q_{\mathrm{cool}}(T) \fullstop
\end{equation}
Defining the equilibrium temperature as $T_0$, we can add a small perturbation
$\delta T$ and Taylor-expand Eq.~(\ref{eq-timedep-cooling-eq}) to first order:
\begin{equation}\label{eq-timedep-cooling-dT}
  \begin{split}
  \frac{e_{\mathrm{th,g}}}{T_0}\, \frac{d\delta T}{dt}\; \simeq &-
  \left(\frac{\partial q_{\mathrm{cool}}}{\partial T}\right)\delta T \comma
  \end{split}
\end{equation}
where we used $e_{\mathrm{th,g}}\propto T$ (Eq.~\ref{eq-e-th-gas}) and the
condition for thermal equilibrium
($q_{\mathrm{cool}}(T_0)=q_{\mathrm{heat}}$). The cooling rate, in the
optically thin limit, is proportional to
$q_{\mathrm{cool}}=q_{\mathrm{cool,\smalldust}}\propto \kappa_P(T)\;T^4$. If the
opacity follows a powerlaw $\kappa_P\propto T^b$, which in our case (see appendix
\ref{app-planck-mean-opacity}) would be $b=1.7$, then we can write
\begin{equation}
  \left(\frac{\partial q_{\mathrm{cool,\smalldust}}}{\partial T}\right)_{T=T_0}
  = (4+b)\; \frac{q_{\mathrm{cool,\smalldust}}(T_0)}{T_0} \fullstop
\end{equation}
Eq.~(\ref{eq-timedep-cooling-dT}) can then be written as
\begin{equation}
  \frac{d\delta T}{dt}\; \simeq - \frac{\delta T}{t^{\mathrm{rad}}_{\mathrm{relax,thin}}} \comma
\end{equation}
with the thermal relaxation time given by
\begin{equation}\label{eq-trelax-thin}
t^{\mathrm{rad}}_{\mathrm{relax,thin}} = \frac{1}{4+b}\;t^{\mathrm{rad}}_{\mathrm{cool,thin}} \comma
\end{equation}
where $t^{\mathrm{rad}}_{\mathrm{cool,thin}}$ is given by
Eq.~(\ref{eq-cooling-time-rad-thin}).}

\section{Thermal relaxation time scale: The optically thick case}
\label{app-thermal-relaxation-time-thick}
\revised{For the optically thick case
we follow \citet{2015ApJ...811...17L}, who derived the relaxation time scale of
a spatial radial temperature fluctuation $\delta T(r,t)\propto e^{ik_xx}$ with
$x=r-r_0$ for the radius $r_0$ where the stability analysis is done, and $k_x$
is the spatial frequency of the wavelike fluctuation. The disk is assumed to be
completely optically thick in vertical direction, and the radial radiative
diffusion between the cooler and warmer part of the wave is assumed to dominate
over the vertical radiative cooling. \citet{2015ApJ...811...17L} show that in
that case, the relaxation time is given by
\begin{equation}\label{eq-trelax-thick}
t^{\mathrm{rad}}_{\mathrm{relax,thick}} = \frac{1}{\eta k_x^2} \comma
\end{equation}
with
\begin{equation}
\eta = \frac{16 \sigma_{SB}T^3}{3\kappa_R(T)\rho_{\smalldust}\rho_gc_{V,g}} \fullstop
\end{equation}
The strongest optical depth effect (i.e., the largest value of
$t^{\mathrm{rad}}_{\mathrm{relax,thick}}$) is found for the smallest realistic
value of $k_x$, corresponding to the largest realistic length scale of the
VSI mode. We set $k_x=2\pi/h_p$ as an estimate of this largest length
scale. The optically thick and thin limits can be combined by adding the
time scales together:
\begin{equation}\label{eq-trelax-both}
t^{\mathrm{rad}}_{\mathrm{relax}}
=t^{\mathrm{rad}}_{\mathrm{relax,thin}}+t^{\mathrm{rad}}_{\mathrm{relax,thick}} \fullstop
\end{equation}
}

\section{Thermal coupling of dust and gas}
\label{app-thermal-coupling-dust-gas}
\revised{The rate of heat transfer between dust and gas is computed by assuming
  perfect thermal accomodation of a gas molecule after it hits a dust
  particle. This leads to a rate of heat transfer per unit surface area of a
  dust particle of}
\begin{equation}
  \tilde q_{\mathrm{dg}} = \rho_g\,\bar C_H\,(T_g-T_{\smalldust}) \comma
\end{equation}
where $T_g$ is the gas temperature, $T_{\smalldust}$ the temperature of the dust particle,
$\rho_g$ the gas density, and $\bar C_H$ the heat exchange coefficient given
by
\begin{equation}
    \bar C_H  = \frac{1}{\gamma-1}\,\sqrt{\frac{k_BT_g}{2\pi \mu m_u}} \, \frac{k_B}{\mu m_u} \fullstop
\end{equation}
The rate of heat transfer per unit volume of the disk is then
\begin{equation}
  q_{\mathrm{dg}}(T_{\smalldust},T_g) = \rho_d\,\rho_g\,\frac{s_d}{m_d}\,\bar C_H\,(T_g-T_{\smalldust}) \comma
\end{equation}
with $s_d=4\pi a^2$ is the surface area of a spherical dust grain of radius
$a$, \revised{which we set to $a=0.1\,\mu$m,}
and $m_d=\tfrac{4\pi}{3}\rho_s a^3$ is its mass, where we set
$\rho_s=1.48\,\mathrm{g/cm}^3$ (see Appendix \ref{app-dust-opacity-model}).
To compute the time scale for heat exchange between the gas and the dust,
relative to the heat content of the gas, we compute
\begin{equation}\label{eq-tdustgas}
t_{\mathrm{dg}} = \frac{e_{\mathrm{th,g}}(T_g)}{q_{\mathrm{dg}}(0,T_g)} \comma
\end{equation}
where $e_{\mathrm{th,g}}(T_g)$ is the thermal energy in the gas, given by
Eq.~(\ref{eq-e-th-gas}). In this equation we set $T_{\smalldust}\rightarrow 0$ because we
are computing a time scale (which, for $T_{\smalldust}=T_g$ would be infinite), not the
actual heat exchange. 

\revised{
We can plot the ratio of $t_{dg}/t^{\mathrm{rad}}_{\mathrm{relax,thin}}$, where
$t^{\mathrm{rad}}_{\mathrm{relax,thin}}$ is given by
Eq.~(\ref{eq-trelax-thin}). The result is shown in Fig.~\ref{fig-tdg-vs-tc} for
the fiducial model, the 10$\times$ less massive disk, and the Herbig Ae star model.}

\begin{figure}
  \centerline{\includegraphics[width=0.5\textwidth]{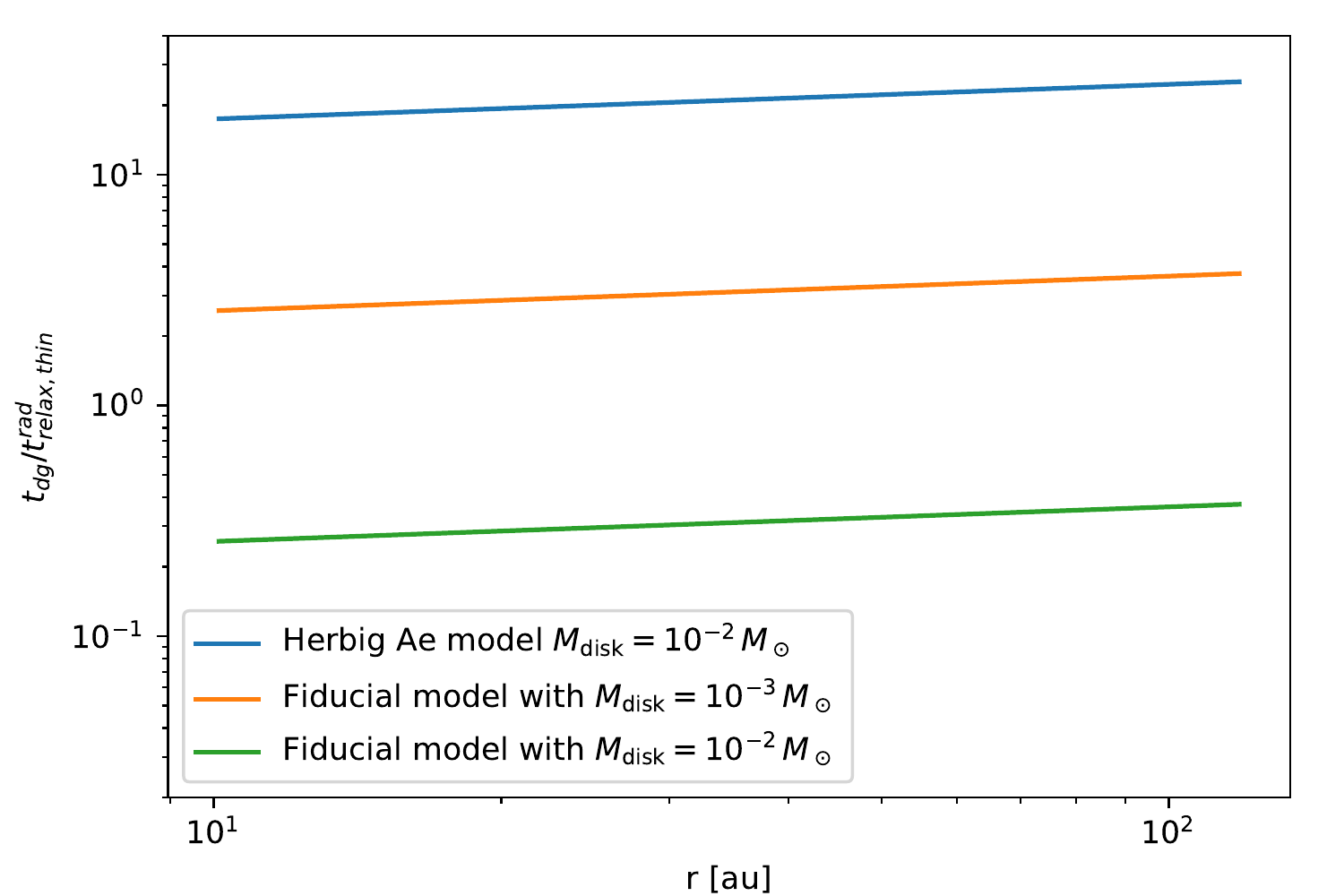}}
  \caption{\label{fig-tdg-vs-tc} Ratio of the dust-gas thermal coupling time
    scale (Eq.~\ref{eq-tdustgas}) to the optically thin radiative relaxation time scale
    (Eq.~\ref{eq-trelax-thin}) at the midplane of the fiducial disk as a
    function of radial coordinate in the disk, for a small-grain size of
    $0.1\,\mu$m. The three curves are the three models presented in this paper.}
\end{figure}

\revised{For the fiducial model with $M_{\mathrm{disk}}=10^{-2}M_\odot$, the value of
$t_{dg}/t^{\mathrm{rad}}_{\mathrm{relax,thin}}$ is consistently below unity, meaning that the
dust-gas thermal coupling is not the limiting factor of the cooling rate. However,
for the less massive disk, the dust-gas coupling time scale starts to dominate, meaning
that for such low mass disks the dust-gas coupling makes it easier to stabilize the
disk against the VSI. For the Herbig Ae star model, the dust-gas coupling time scale
dominates over the thermal relaxation time in the optically thin limit, but
(as seen in Fig.~\ref{fig-tcool-vsi-herbigae}), when optical depth effects start
to play a role, the thermal relaxation time again dominates over the dust-gas
thermal coupling time.}

\revised{The inclusion of the dust-gas thermal coupling time is particularly important
for the Herbig Ae star case. Without including it, the curves in
Fig.~\ref{fig-tcool-vsi-herbigae} would be much lower (more favorable for
the VSI).}

\section{Optical appearance of a disk with strong depletion of small grains}
\label{sec-opt-appearance-depletion}

\revised{A concern with the depletion of small grains to inhibit the VSI is that
  too much depletion would affect the appearance of the disk as seen in
  scattered light in the optical and near-infrared. At these wavelengths the
  stellar light scatters off the small dust grains in the surface layers of the
  disk, and the disk appears on the sky with a distinct ``hamburger'' shape,
  with light seen on two sides of a dark lane. If, however, the small grains are
  too strongly depleted, then this shape becomes flatter or disappears
  altogether. The question is: does a depletion factor of 0.1 or 0.01, required
  to inhibit the VSI, leave enough small grains in the disk to appear on the sky
  with this ``hamburger'' shape?}

\begin{figure*}
  \centerline{\includegraphics[width=0.95\textwidth]{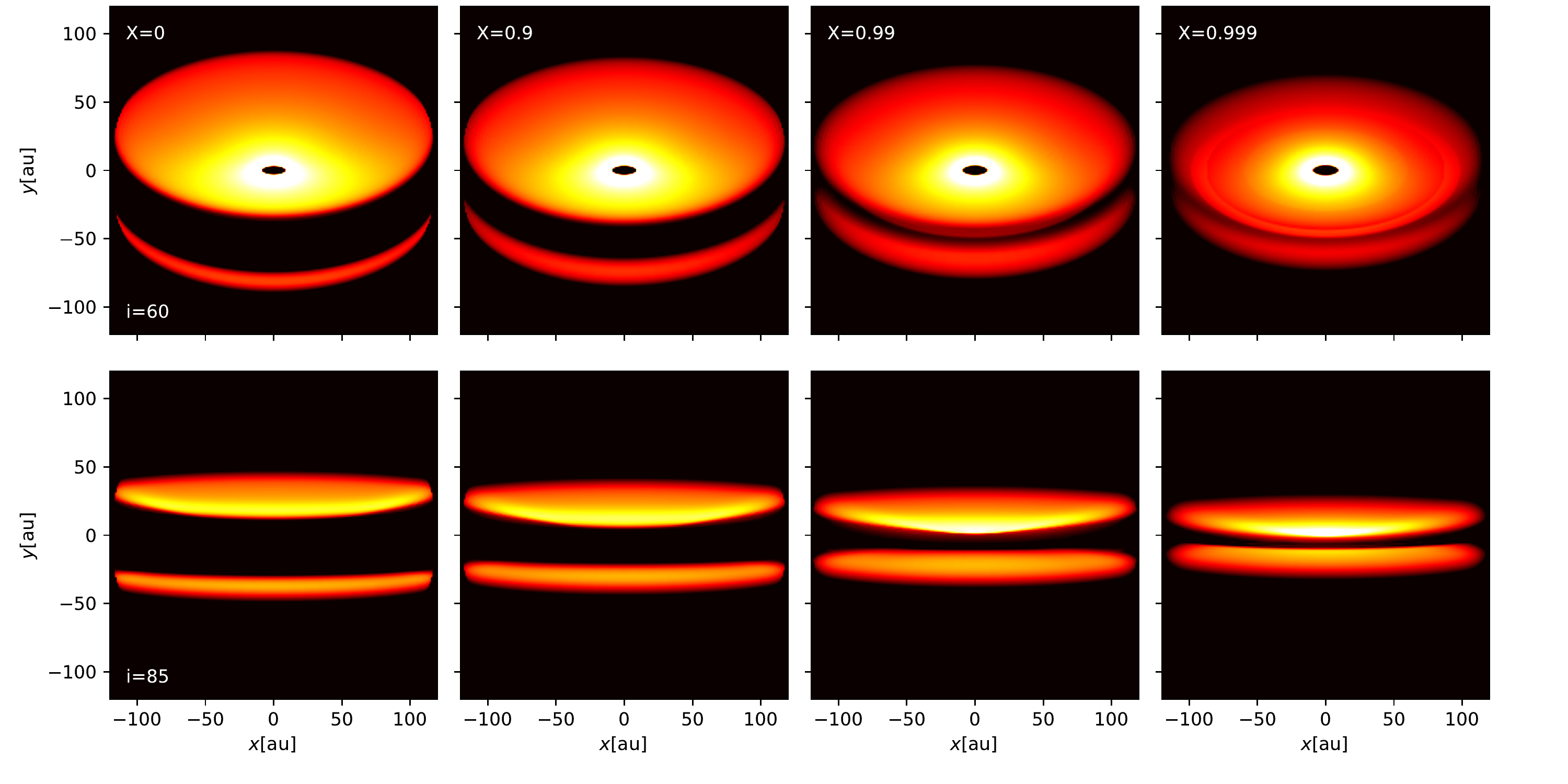}}
  \caption{\label{fig-depletion-appearance}Appearance of the fiducial disk
    model at $\lambda=0.8\,\mu$m for four values of the coagulation parameter
    $\coag=0$, 0.9, 0.99 and 0.999 (left to right), corresponding to a
    small-grain-depletion factor of 1, 0.1, 0.01 and 0.001. Top panels:
    inclination of $i=60$. Bottom panels: inclination of $i=85$. All images have
    the same logarithmic color scale spanning 3 factors of ten.}
\end{figure*}

\revised{To find out, we run four radiative transfer models with RADMC-3D, all
  based on the fiducial disk model of appendix \ref{app-fiducial-disk-model},
  but with $\coag=0$, 0.9, 0.99 and 0.999 (i.e., small-grain depletion factors
  of 1, 0.1, 0.01 and 0.001), and render images of them at $\lambda=0.8\,\mu$m
  and inclinations $i=60$ and $i=85$. The star is removed from the images.  Both
  the small grains and the big grains are included in the model. For the big
  grain opacity at this wavelength, the extreme forward-peaked part of the
  scattering phase function is ``chopped'' within 10 degrees. This is a way to
  keep the Monte Carlo radiative transfer well-behaved without the need for an
  extremely high number of photon packages (see the manuals of RADMC-3D and of
  optool).  The results are shown in Fig.~\ref{fig-depletion-appearance}.}

\revised{It is clear that for all depletion factors the disk still appears in
  its characteristic shape with top and bottom bright layers separated by a dark
  lane. For stronger depletion of small grains the dark lane becomes narrower,
  as expected. For $\coag=0.99$ and $\coag=0.999$ the disk becomes vertically
  sufficiently optically thin that some of the starlight, after being scattered
  off the surface dust, reaches the midplane layer of large dust grains, which
  is visible in the images, albeit at very low brightness.}

\revised{Of course, this also depends on the disk mass. And not all disks look
  alike. It therefore requires a case-by-case analysis. But overall we can
  conclude that the explanation of inhibiting the VSI by grain growth is not
  inconsistent with the typical appearance of protoplanetary disks at visual and
  near-infrared wavelengths.}

\section{Comparison to a simple settling-mixing equilibrium}
\label{sec-settling-mixing}

\begin{figure*}
  \centerline{\includegraphics[width=0.95\textwidth]{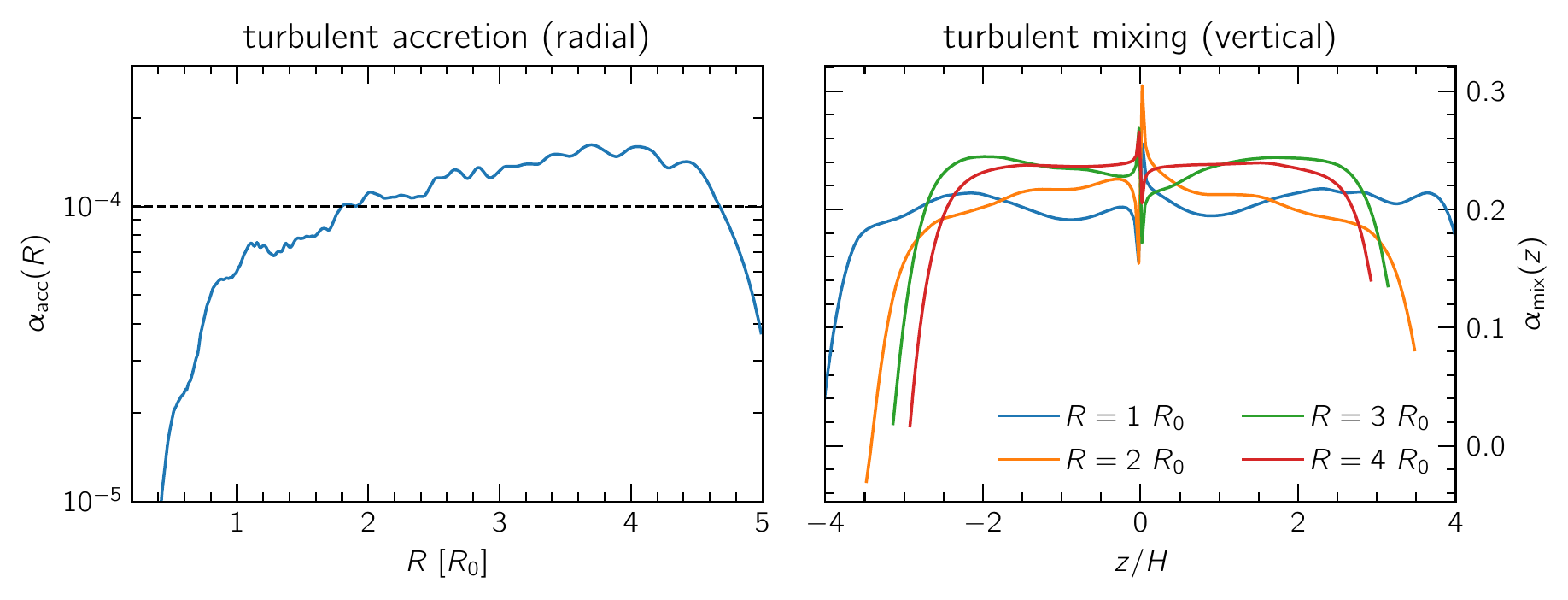}}
  \caption{\label{fig-vsi-alphas}Measured turbulent radial-azimuthal
    viscosity and vertical mixing $\alpha$ coefficient, using the method of
    \citet{2017A&A...599L...6S}. Left: The viscous $\alpha$. Right: The vertical
    mixing $\alpha$ as a function of $z$ in units of the pressure scale height
    at four radii (in units of $r_0$).}
\end{figure*}

It is instructive to compare the results of the particle motions to a simple vertical
settling-mixing calculation such as those of \citet{1995Icar..114..237D},
\citet{2004A&A...421.1075D} and \citet{2009A&A...496..597F}. The measured radial
and vertical $\alpha$ of our model, using the method of
\citet{2017A&A...599L...6S}, are shown in Fig.~\ref{fig-vsi-alphas}. It is
evident that the turbulent motions of the VSI are highly anisotropic. The
vertical turbulence parameter $\alpha_{\mathrm{mix}}$ is about 1700 times larger
than the radial turbulence parameter $\alpha_{\mathrm{acc}}$.  For the vertical
mixing-settling calculation we use $\alpha_{\mathrm{mix}}$. According to
\citet{2010A&A...513A..79B}, their Eq.~(51), the geometric thickness
$H_{\bigdust}$ of the big grain dust layer can then be estimated as
\begin{equation}\label{eq-dust-layer-height}
H_{\bigdust} \simeq h_p \min\left(1,\sqrt{\frac{\alpha_{\mathrm{mix}}}{\min(\mathrm{St},1/2)(1+\mathrm{St}^2)}}\right) \comma
\end{equation}
where the outer min() operator is because $H_{\bigdust}$ is defined such that it
cannot exceed $h_p$. For $\alpha_{\mathrm{mix}}\simeq 0.17$ and $\mathrm{St}\ll
1$, Eq.~(\ref{eq-dust-layer-height}) reduces to $H_{\bigdust} \simeq
h_p\min(1,\alpha_{\mathrm{mix}}/\mathrm{St})$, and the dust will be almost as
vertically extended as the gas ($H_{\bigdust}\simeq h_p$), which is indeed what
we find. For $\mathrm{St}\gg 1$, Eq.~\ref{eq-dust-layer-height} reduces to
$H_{\bigdust} \simeq h_p\sqrt{2\alpha_{\mathrm{mix}}}/\mathrm{St}$. This matches
the conveyor-belt estimate from Section \ref{sec-conveyor-belt-estimate} if
we set $|v_{z,\mathrm{VSI}}|/c_s=\sqrt{2\alpha_{\mathrm{mix}}}$. However, from
Fig.~\ref{fig-vsi-vz-time} we know that the typical values of
$|v_{z,\mathrm{VSI}}|/c_s$ are of the order of 0.1, while the measured value
of $\sqrt{2\alpha_{\mathrm{mix}}}$ from the simulation is about 0.58. It
shows the limitations of these simple estimates.

\end{document}